\documentclass[preprint,prb,amsmath,aps,subeqn]{revtex4}
\usepackage{graphicx}
\usepackage{epsfig}
\usepackage[cp1251]{inputenc}
\usepackage[english]{babel}
\usepackage{amsmath}
\usepackage{amssymb}
\usepackage{amstext}
\usepackage{color}
\usepackage[toc,page]{appendix}
\usepackage{comment}

\begin{document}

\title{Spatially dispersive dynamical response of hot carriers in doped graphene}

\author{S. M. Kukhtaruk,$^{1}$ V. A. Kochelap,$^{1}$ V. N. Sokolov,$^{1}$ and K. W. Kim$^{2}$}
\affiliation{$^{1}$Department of Theoretical Physics, V. Lashkaryov Institute of Semiconductor Physics, NASU,
 Pr. Nauki 41, Kiev 03028, Ukraine}
 \affiliation{$^{2}$Department of Electrical and Computer Engineering,
North Carolina State University, Raleigh, NC 27695-7911, USA}

\begin{abstract}
We study theoretically wave-vector and frequency dispersion of the complex dynamic conductivity tensor (DCT), $\sigma_{lm}(\mathbf{k}, \omega)$,  of doped monolayer graphene under a strong dc electric field. For a general analysis, we consider the weak ac field of arbitrary configuration given by two independent vectors, the ac field polarization and the wave vector $\mathbf{k}$. The high-field transport and linear response to the ac field are described on the base of the Boltzmann kinetic equation. We show that the real part of DCT, calculated in the collisionless regime, is not zero due to dissipation of the ac wave, whose energy is absorbed by the resonant Dirac quasiparticles effectively interacting with the wave. The role of the kinematic resonance at $\omega = v_F |{\bf k}|$ ($v_{F}$ is the Fermi velocity) is studied in detail taking into account deviation from the linear energy spectrum and screening by the charge carriers. The isopower-density curves and distributions of angle between the ac current density and field vectors are presented as a map which provides clear graphic representation of the DCT anisotropy. Also, the map shows certain ac field configurations corresponding to a negative power density, thereby it indicates regions of terahertz frequency for possible electrical (drift) instability in the graphene system.

\end{abstract}

\maketitle
\section{Introduction}\label{B-1}

One of fundamental aspects of the graphene physics is the electronic band dispersion which is linear $\varepsilon({\bf p}) =\pm v_{F} p$ in the vicinity of the Dirac points for a wide region of energy $\varepsilon$ ($<$ 1 eV);~\cite{Wallace,Sarma_rev} here $v_{F} \simeq 10^{8}$ cm/s is the Fermi velocity in graphene, ${\bf p}$ is the electron momentum, and $p = |{\bf p}|$. Hence, the electro-physical properties essentially determined by the energy band structure will be different for graphene and conventional semiconductors with the standard energy band (Ge, Si, GaAs, etc).~\cite{Neto}

A lot of attention has been paid to the investigation of the electrical conductivity in graphene both experimentally and theoretically.
A special focus is given to the investigation of electron dynamic response to electrical perturbations which can be varying in both space and time.~\cite{Falkovsky_2008,Lloyd,Horng,Lin,Mak,Falkovsky_2007,Gusynin,Li,Hao,Lovat}
Generally, spatial dispersion for arbitrary wave vectors should be included in the description of the dynamic conductivity which, in this case, due to the existence of preferential direction given by the wave vector ${\bf k}$ is a tensorial quantity, $\sigma_{lm}(\mathbf{k}, \omega)$, the dynamic conductivity tensor (DCT) .
The detailed knowledge of wave-vector dispersion of DCT is critically important for use in tunable THz plasmonic and metamaterial nanodevices based on excitation modes in different graphene structures.\cite{Luo,Wagner} The examples are patterned structures such as periodic micro-ribbon arrays,~\cite{Rana,Ju} antidots arrays,~\cite{Nikitin} grating-gated graphene-based HEMTs,~\cite{Esfahani} and other semiconductor plasmonic systems.~\cite{OurJAP_2011,OurUkrJP_2012,Korotyeyev} Also, a scheme which avoids patterning has been proposed for coupling light into graphene plasmons by forming a tunable optical grating with electrically generated surface acoustic waves.~\cite{Schiefele} Recently, the novel nanoscope techniques have been developed where electric fields (variable in time and space) can be excited at the length scale of the order of tens of nanometers, as well as the methods which allow to explore the excited fields and plasmons on these time and length scales, including ultrafast resonance of Dirac plasmons in graphene and similar systems.~\cite{Hillenbrand,Fei1,Huber,Fei2,Chen,Gonsales, Mishchenko, OurPRB_2015} Another important feature associated with the wave-vector dispersion in such systems is that the wave vector ${\bf k}$ and vector of the locally excited field are not parallel as it takes place, for example, in the case of electrostatic perturbation, where they are always collinear according to the Poisson equation.

Since the discovery of graphene,~\cite{Novoselov} the behavior of Dirac quasiparticles in external electric fields is the subject of intense theoretical and experimental studies. Such studies are stimulated by both the exceptional electronic properties of graphene~\cite{Neto} and the perspective of various ultrahigh-speed applications in future integrated-circuit technology,~\cite{Schwierz} high-frequency electronics including terahertz (THz) frequencies,~\cite{Lin,Lloyd,Liu} and optoelectronics.~\cite{Mueller} In early research, most transport studies in graphene have been carried out under low-bias conditions to probe the intrinsic electronic properties, although high-field behavior of the nonequilibrium carriers is more appropriate for practical graphene-based device operation.~\cite{Tahy,DaSilva,Perebeinos, Liao,Bae_1, Bae_2} The wave-vector $\mathbf{k}$ and frequency $\omega$ dispersion of the graphene DCT $\sigma_{lm}(\mathbf{k}, \omega)$ was considered in the collisionless regime~\cite{Falkovsky_2007} and in the relaxation-time approximation~\cite{Lovat} for a weak electric field, which did not break the equilibrium distribution of Dirac quasiparticles over energy in the bands.
In the high-field limit, the charged carrier transport is described, in particular, in terms of the hot electrons and/or holes in graphene.~\cite{DaSilva,Bistritzer,Svintsov_1, Svintsov_2,Serov}


In a general case of arbitrary wave vector $\mathbf{k}$ and frequency $\omega$, the dispersion of DCT becomes important for such $\mathbf{k}$ and $\omega$ which obey the inequalities $kv\tau_p \gtrsim 1$, $\omega \tau_p  \gtrsim 1$, where ${\bf v} = \partial \varepsilon/\partial {\bf p}$ and $\tau_p$ are the carrier band velocity and the momentum relaxation time, respectively; $\textstyle{v=|{\bf v}|}$ and $k = |\mathbf{k}|$. The collisionless limit corresponds to the conditions
\begin{equation} \label{collisionless}
 kv\tau_p \gg 1, \;\; \omega \tau_p \gg 1 \,,
\end{equation}
which imply many oscillations of the ac field on the space and time scales given by the carrier mean free path $l_p = v\tau_p$ and the characteristic time $\tau_p$, respectively. These allow to neglect a small collision integral, compared to the rest of terms in the Boltzmann kinetic equation, and to calculate the DCT, $\sigma_{lm}(\mathbf{k}, \omega)$, in the straightforward manner. In this regime, the space and time dispersion results from the fact that {\it free movement} of the carriers is affected by all values of the ac field along the trajectories of movement rather than its local and instantaneous value.

In this paper, we investigate theoretically the space and time dispersion of DCT of Dirac quasiparticles in the doped single-layer graphene system under a high dc electric field. For generality, we consider the perturbing ac field to be of arbitrary configuration, that is the electric field vector and the wave vector ${\bf k}$
are treated as the two independent vectors. The theoretical model is based on the semiclassical approach which is valid for $k < k_F$ and $\omega < \varepsilon_F/\hbar$, where $k_F$ and $\varepsilon_F$ are the Fermi wave vector and energy, respectively. The latter inequality allows us to exclude the interband transitions between the valence and conduction bands which assumes frequency $\omega < 2 \varepsilon_F/\hbar$.

The specific behavior of real, $\sigma^{\prime}_{lm}(\mathbf{k}, \omega)$, and imaginary, $\sigma^{\prime \prime}_{lm}(\mathbf{k}, \omega)$, parts of the complex DCT $\sigma_{lm}(\mathbf{k}, \omega) = \sigma^{\prime}_{lm}(\mathbf{k}, \omega) +i \sigma^{\prime \prime}_{lm}(\mathbf{k}, \omega)$ is tightly connected to the well-known kinematic resonance in the collisionless plasma.~\cite{Landau} At the resonance $\omega = \mathbf{k\cdot v}$, the phase velocity of the wave in the direction of propagation $\omega/k$ is equal to the carrier velocity $v$. For graphene, due to the linear energy spectrum, the resonance is determined by {\it a linear} dependence of frequency on wave vector, $\omega = v_F k$, which is to some extent similar to the Dirac cone $\varepsilon/\hbar =  v_{F}q$; here ${\bf q}$ is the Dirac quasiparticle wave vector and $q = |{\bf q}|$. We will show that for high frequencies $\omega > v_F k$ [fast waves, $(\omega/k) > v_F$]  the real part of DCT becomes equal to zero. In the region of low frequencies $\omega < v_F k$ [slow waves, $(\omega/k) < v_F$], the space dispersion is large and the real part of DCT takes definite nonzero values. We note that such specific dispersion of $\sigma^{\prime}_{lm}(\mathbf{k}, \omega)$ is due to the intrinsic property of the graphene energy spectrum and totaly different from standard 2D electron systems with the parabolic energy spectrum.

This paper is organized as follows. We introduce our model and describe the theoretical approach in Sec.~II and use them to study the real part (Sec.~III) and the imaginary part (Sec.~IV) of the complex DCT. Also in Sec.~III, we provide a graphic representation of the DCT anisotropy in the form of a map for isopower-density curves and distributions of angle between the ac current density and electric field vectors. In Secs.~V and VI, we discuss the role of the energy band nonlinearity and screening by the charge carriers, respectively. We present conclusions in Sec.~VII. In the Appendix, we describe details of the evaluation of integrals in the expressions for the DCT.

\section{Formulation of the model}\label{B-2}

In our study, we consider doped graphene systems in which the electron (hole) density is controlled by an external gate voltage and/or chemical doping.~\cite{Neto} As far as the graphene spectrum is symmetric relative to Dirac electron and hole quasiparticles, we will refer, for definiteness, to the electron-doped graphene with the electron density $n_0$. We assume the Fermi level is well above the Dirac point for such doping which allows us to neglect the contribution of holes. For the calculation of DCT, we consider a graphene monolayer sheet lying in the $XOY$ plane ($z = 0$), driven by a strong dc electric field ${\bf E}_0$ and a weak ac electric field ${\bf E}_1({\bf r},t) = {\bf E}_1 e^{i({\bf k}\cdot{\bf r} - \omega t)}$ with $|{\bf E}_1| \ll |{\bf E}_0|$; ${\bf r} = (x,y)$ is the in-plane coordinate vector and $t$ is the time. In the chosen coordinate system, the wave vector $\textstyle{{\bf k}=(k, 0)}$ has only one nonzero component $k_x = k$ (Fig. 1). The complex amplitude of the ac field (and other perturbation quantities) is frequency and wave-vector dependent, ${\bf E}_1 = {\bf E}_1^{\omega{\bf k}}$. For shortness of notations, we will omit these subindices throughout the text except Sec. VI. For linearization purposes, we use the usual plane-wave ($\omega, {\bf k}$) representation. However, where it is necessary [see Eq.~(\ref{sigma lm})], we will utilize the substitution $\omega \rightarrow \omega + i\delta$ ($\delta \rightarrow +0$) which corresponds to a slow turning on the ac field at $t = -\infty$.~\cite{Landau} Notice that the weak ac electric field can correspond to the electric field of a propagating electromagnetic wave, the effective field of electrostatic or deformation potentials, or a weak inhomogeneity of different origin. In general, it can depend on the $z$-coordinate as well, ${\bf E}_1({\bf r},z,t)$. However, in the simplest approach when the graphene is modeled as an infinitesimally thin layer in the $z$ direction (a delta-layer), the main expressions derived below contain the field value at $z = 0$. We designate the in-plane value of the field as ${\bf E}_1({\bf r},t) = {\bf E}_1({\bf r},z=0,t)$.

\begin{figure}\label{Fig1}
\begin{center}
\includegraphics[width=8.5cm]{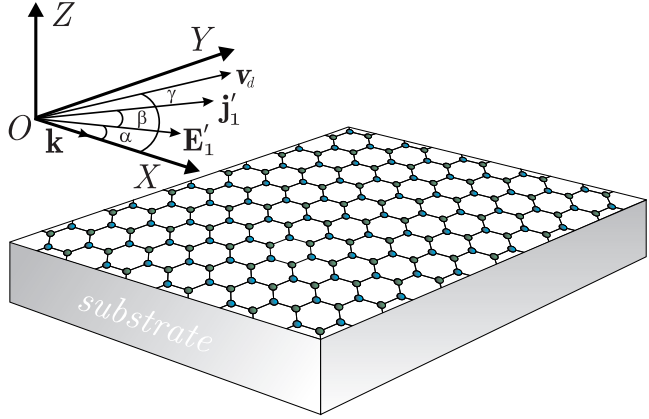}
\caption{Geometry of the problem. Graphene single layer (on a substrate) and coordinate system show in-plane basic vectors and characteristic angles appearing in the problem: electron drift velocity ${\bold v}_d$, amplitudes of ac electric field ${\bold E}^{\prime}_1$ and current density ${\bold j}^{\prime}_1$ with wave vector ${\bold k}$ and frequency $\omega$. The angles $\alpha, \beta, \gamma$ are marked by arcs.}
\end{center}
\end{figure}

To describe the linear response of nonequilibrium Dirac quasiparticles (electrons) to the ac electric field, we result from the Boltzmann equation
\begin{equation}\label{kinetic}
 \frac{\partial f}{\partial t} + v_{F} \frac{{\bf p}}{p} \frac{\partial f}{\partial {\bf r}} -e{\bf E} \frac{\partial f}{\partial {\bf p}} = I \{f \} \,,
\end{equation}
where ${\bf E} = {\bf E}_0 + {\bf E}_1({\bf r},t)$ is the lateral electric field, $e = |e|$ is absolute value of the electron charge, and $I = I_{eL} + I_{ee}$ is the collision integral which includes different scattering mechanisms such as electron-phonon and impurity scattering ($I_{eL}$) and electron-electron scattering ($I_{ee}$). The distribution function can be represented as $f({\bf r},{\bf p},t) = f_0({\bf p}) + f_1({\bf r},{\bf p},t)$, where $f_0({\bf p})$ is time-independent and space-uniform distribution function of the electrons in the dc field ${\bf E}_0$ and $f_1({\bf r},{\bf p},t)$ is a small addition due to the ac field ${\bf E}_1({\bf r},t)$. Letting a harmonic dependence on the coordinate and time for all quantities [e.g., $f_1({\bf r},{\bf p},t) = f_1({\bf p})e^{i({\bf k}\cdot{\bf r} - \omega t)}$, etc] and utilizing the usual procedure of linearization in Eq.~(\ref{kinetic}), we obtain the equations for the steady-state distribution function $f_0({\bf p})$
\begin{equation}\label{kinetic f0}
  e{\bf E}_0 \frac{\partial f_0}{\partial {\bf p}} + I \{f_0 \} = 0
\end{equation}
and the time-dependent addition $f_1({\bf r},{\bf p},t)$
\begin{equation}\label{kinetic f1}
 \frac{\partial f_1}{\partial t} + v_{F} \frac{{\bf p}}{p} \frac{\partial f_1}{\partial {\bf r}} -e{\bf E}_1 \frac{\partial f_0}{\partial {\bf p}} -e{\bf E}_0 \frac{\partial f_1}{\partial {\bf p}} = I \{f_0,f_1 \} \,.
\end{equation}
For large electron densities, the $e-e$ scattering time appears to be shorter than all other scattering times, so that the leading term in Eq.~(\ref{kinetic f0}) is determined by the $e-e$ collision integral. In this regime, rapid $e-e$ collisions establish a displaced Fermi-Dirac distribution function
\begin{equation}\label{f0}
 f_{0} ({\bf p})=\left[\exp{\left(\frac{v_F p - \mathbf{v}_d \mathbf{p} - \varepsilon_F }{k_B T_e}\right)}  +1\right]^{-1} \,,
\end{equation}
which satisfies the equation $I_{ee}\{f_0\} = 0$ and thus can be regarded as an approximate solution to Eq.~(\ref{kinetic f0}).~\cite{Gantmakher} In Eq.~(\ref{f0}), ${\bf v}_d$ is the electron drift velocity which is collinear with the dc field ${\bf E}_0$, and $k_B$ is the Boltzmann constant. The three parameters, the Fermi energy, $\varepsilon_F$, the average electron drift velocity, ${\bf v}_d$, and the electron temperature, $T_e$, depend on ${\bf E}_0$ and describe the steady state of the nonequilibrium graphene system in the dc electric field ${\bf E}_0$. These parameters are found from the set of coupled equations which are pertinent moments of the kinetic equation (\ref{kinetic f0}) and represent balance equations for the carrier density, the electron momentum, and the electron energy.~\cite{DaSilva,Bistritzer,Svintsov_1, Svintsov_2}
Integrating Eq.~(\ref{f0}) over $d\Gamma_p = gd^2p/(2\pi\hbar)^2$, where $g = g_sg_v = 4$ is the electron spin ($g_s = 2$) and valley ($g_v = 2$) degeneracy factor, we obtain
\begin{equation}\label{epsilon F}
  \frac{g}{(2\pi\hbar)^2} \int d^2p f_{0}({\bf p}) = n_0  \,.
\end{equation}
For strongly degenerate electrons, the integral can be performed analytically and the asymptotic relationship can be found for the Fermi energy $\varepsilon_F = \hbar k_F v_F (1 - v_d^2/v_F^2)^{3/4}$, where $k_F= \sqrt{\pi n_0}$ is the Fermi wave vector in the absence of the dc electric field.
To avoid considerable decrease in $\varepsilon_F$ with increasing drift velocity, we exclude from our consideration such values of $v_d=|{\bf v}_d|$ (the dc field strength $E_0$) for which the relativistic factor $v_d/v_F$ approaches unity. In the dc transport, the momentum and energy gained by electrons from the electric field are balanced by losses due to electron scattering on the impurities and phonons. The results of numerical calculations of the field dependencies of ${\bf v}_d = \mathbf{v}_d(\mathbf{E}_0)$ and $T_e = T_e(\mathbf{E}_0)$ can be found elsewhere (see, for example, Ref.~\onlinecite{DaSilva}). In this work, we utilize $\mathbf{v}_d$ and $T_e$ as the known model parameters for the analysis of the DCT.

An approximate analytical solution to Eq.~(\ref{kinetic f1}) can be found in the collisionless regime by neglecting the collision integral $I \{f_0,f_1 \}$. This approach is valid for the range of frequencies and wave vectors within the criteria given in  (\ref{collisionless}). Also, we omit the last term on the left-hand side of Eq.~(\ref{kinetic f1}) to obtain
\begin{equation}\label{f1}
  f_1({\bf p}) = \frac{ie{\bf E}_1}{\omega - v_F {\bf k}\cdot{\bf p}/p} \frac{\partial f_0}{\partial {\bf p}}   \,.
\end{equation}
The latter assumption imposes restrictions on the dc field strength according to the following inequalities $(eE_0/\omega) \ll \bar{p}$, $(eE_0/k) \ll v_F\bar{p}$. Physically, they  mean that the momentum and energy acquired by an electron on the time and space scales of the order of the ac field period ($\sim 1/\omega$) and wavelength ($\sim 1/k$) are much less than the average momentum $\bar{p}$ and energy $v_F\bar{p}$, respectively.

The ac current density ${\bf j}_1({\bf r},t) = {\bf j}_1 e^{i({\bf k}\cdot{\bf r} - \omega t)}$ is related to the ac field ${\bf E}_1({\bf r},t)$ through the DCT ${\bf j}_1 = \hat{\sigma} {\bf E}_1$. Substituting $f_1({\bf p})$ from Eq.~(\ref{f1}) into the current density
\begin{equation}\label{j1}
  {\bf j}_1 = - \frac{eg}{(2\pi\hbar)^2} \int d^2p \, v_F \frac{{\bf p}}{p} f_1({\bf p})  \,,
\end{equation}
we obtain the DCT as
\begin{equation}\label{sigma lm}
  \sigma_{lm} = - i \frac{e^2 gv_F}{(2\pi\hbar)^2} \int \frac{d^2 p}{\omega - v_F {\bf k}\cdot{\bf p}/p + i\delta}
  \frac{p_l}{p} \frac{\partial f_0}{\partial p_m}  \,.
\end{equation}
Under the integral, an infinitesimally small imaginary $i\delta$ in the denominator corresponds to a slow turning on the ac field at $t = -\infty$ and results in the regularization of the integral at $\delta \rightarrow +0$.~\cite{Landau} As it can be seen from  Eq.~(\ref{sigma lm}), there is no singularity in the integrand for frequency $\omega > v_F k$. This is the key difference of the pole $\omega = v_F {\bf k}\cdot{\bf p}/p$ which takes place for the graphene from a similar pole $\omega = {\bf k}\cdot{\bf p}/m$ characteristic for electron systems with the standard (quadratic) energy spectrum $\varepsilon = p^2/2m$. Thus, the equation $\omega = v_F {\bf k}\cdot{\bf p}/p$ has solutions only at $\omega < v_F k$, in contrast to the equation $\omega = {\bf k}\cdot{\bf p}/m$ which has solutions at any $\omega$ and $k$. This results in that the real part of the DCT is not zero only for frequencies $\omega < v_F k$.

The dynamic conductivity tensor $\sigma_{lm}$ is a complex tensor of the second order. It is given by the eight functions which depend on the six parameters: $\omega, k$, and $\gamma$ which are the propagating wave characteristics, and $n_0, v_d$, and $T_e$ which are the dc transport characteristics.
With the steady-state distribution function (\ref{f0}), the integrals in Eq.~\eqref{sigma lm} can be performed analytically. In particular, for $v_d = 0$, the function $(\ref{f0})$ represents the so-called {\it electron temperature approximation} which is widely used in the theory of the high-field carrier transport.~\cite{Gantmakher} In this case, the off-diagonal components of tensor $\hat{\sigma}$ are zero and obey the evident symmetry relations $\sigma_{xy} = \sigma_{yx} = 0$. The same occurs for $\sin\gamma = 0$, i.e., for the drift velocity $\textstyle{\mathbf{v}_d}$ and the wave vector $\textstyle{\bf k}$ to be collinear (Fig. 1).

Below we analyze the behavior of the real $\sigma^{\prime}_{lm}$ and imaginary $\sigma^{\prime\prime}_{lm}$ parts of $\sigma_{lm}$ separately. First, we address the real part of the DCT.

\section{Real part of dynamic conductivity}

We now turn to explicit evaluation of the integrals in Eq.~\eqref{sigma lm}. In this and in the following Sec. IV, we use the linear energy dispersion for graphene, and later (Sec. V) we take into account deviation from the linear band structure to analyze its influence on the DCT. It is convenient to introduce the dimensionless variables $\Omega = \omega/\omega_0$ and $K = v_Fk/\omega_0$ where the normalization frequency is $\omega_0 = e^2gk_BT_e/(2\pi\hbar^2v_F)$, and the dimensionless parameters $V_d = v_d/v_F$ ($V_d < 1$) and $E_F = \varepsilon_F/k_BT_e$. After substitution of the distribution function (\ref{f0}) into Eq.~\eqref{sigma lm}, the integration therein can be performed analytically utilizing the complex plane and the residue theory method.~\cite{Morse} The calculation details are described in the Appendix where the DCT is given by Eq.~\eqref{sigma exact}. Since the explicit form of the general expression for $\sigma_{lm}$ in Eq.~\eqref{sigma exact} is rather complicated, it is difficult to separate analytically the real and imaginary parts of $\sigma_{lm}$ from this equation. We will utilize Eq.~\eqref{sigma exact} in Sec. IV for the numerical calculation and analysis of the imaginary part of $\sigma_{lm}$.
We notice that it is more simple to separate $\sigma^{\prime}_{lm}$ in Eqs.~\eqref{IntPoAlpha}, where the integration is over the polar angle $\theta$. Indeed, in using the well-known relation $(x + i\delta)^{-1} = {\cal P}(1/x) - i\pi\delta(x)$ [${\cal P}(1/x)$ denotes the principal part of the integral and $\delta(x)$ is the delta function], it is clear that the integral with the delta function is easily evaluated, and the explicit analytical expressions for $\sigma^{\prime}_{lm}$ are obtained
\begin{eqnarray} \label{resigma}
 \sigma^{\prime}_{xx}=  \frac{B\Omega}{\sqrt{K^2-\Omega^2}} \left(\frac{\Omega-V_xK}{A_-^2}+\frac{\Omega-V_xK}{A_+^2}\right),   \\
 \sigma^{\prime}_{yy}= B \left(\frac{\sqrt{K^2-\Omega^2}-V_yK}{A_-^2}+\frac{\sqrt{K^2-\Omega^2}+ V_yK}{A_+^2}\right),  \nonumber    \\
 \sigma^{\prime}_{xy}=  \frac{B\Omega}{\sqrt{K^2-\Omega^2}} \left(\frac{\sqrt{K^2-\Omega^2}- V_yK}{A_-^2}-\frac{\sqrt{K^2-\Omega^2}+ V_yK}{A_+^2}\right),  \nonumber \\
 \sigma^{\prime}_{yx}=  B \left(\frac{\Omega-V_xK}{A_-^2}-\frac{\Omega-V_xK}{A_+^2}\right); \nonumber
\end{eqnarray}
here
\begin{eqnarray} \label{ABCresigma}
 A_{\pm}=K-V_x\Omega\pm V_y\sqrt{K^2-\Omega^2},  \\
 B=\frac{v_F}{2}\ln{(e^{E_F}+1)}\,\theta(K-|\Omega|), \nonumber
\end{eqnarray}
$\textstyle{V_x=V_d\cos{\gamma}}$, $\textstyle{V_y=V_d\sin{\gamma}}$, and $\theta(x)$ is the Heaviside step function. We have verified that the real part of the DCT obtained in such manner coincides with that derived from Eq.~\eqref{sigma exact} in the Appendix.

It follows from Eqs.~\eqref{resigma} that $\sigma^{\prime}_{yy}$ is nonnegative while the other components can take both signs. It is also evident that $\sigma^{\prime}_{lm} = 0$ at $|\Omega| > K$; if the frequency approaches to the resonance $\Omega\rightarrow K$, then $\sigma^{\prime}_{xx}$ and $\sigma^{\prime}_{xy}$ go to infinity  $ \sim (K - \Omega)^{-1/2}$, whereas $\sigma^{\prime}_{yx}$ and $\sigma^{\prime}_{yy}$ approach to zero
$\sim (K - \Omega)^{1/2}$. It is seen that if  $\sin{\gamma} = 0$  or $\textstyle{V_d=0}$, then $V_y= 0$ and  $A_{+} = A_{-}$ so that the real part of DCT becomes symmetrical tensor, with $\sigma^{\prime}_{xy} = \sigma^{\prime}_{yx} =0$. In the analysis of anisotropy properties of $\sigma^{\prime}_{lm}$, it is instructive to consider the following four particular cases: 1. $v_d = 0$; 2. $v_{dx} \neq 0, v_{dy} = 0$; 3. $v_{dx} = v_{dy} \neq 0$; and 4. $v_{dx} = 0, v_{dy} \neq 0$. For numerical calculations, we choose the set of parameters: $K = 1.5$, $V_d = 0.4$, $n_0 =2\times10^{12}$ cm$^{-2}$, and $T_e = 400$ K ($k_BT_e \simeq 34.5$ meV).~\cite{DaSilva} With these parameters, we estimate $\omega_0 \simeq 7.3 \times 10^{13}$~s$^{-1}$, $k_F \simeq 2.5\times 10^6$ cm$^{-1}$, $\varepsilon_F \simeq 153$ meV ($V_d = 0$), and $\varepsilon_F \simeq 131$ meV ($V_d = 0.4$); $k \simeq 1.1 \times10^{6}$ cm$^{-1}, \lambda = 2\pi/k \simeq 57$ nm, and the resonance frequency ($\Omega = K$) $\nu \simeq 17$ THz.

1. {\it In the first case}, $V_d = 0$ so that there is no drift of the electron system as a whole in the dc electric field (the electron temperature $T_e$ can be higher than the lattice temperature). The diagonal components in Eqs.~\eqref{resigma} are expressed as
\begin{eqnarray} \label{resigmak0v0}
 \sigma^{\prime}_{xx} = v_F\ln(e^{E_F}+1) \frac{\Omega^2}{K^2 \sqrt{K^2-\Omega^2}} \,\theta(K-|\Omega|),  \\
 \sigma^{\prime}_{yy} = v_F\ln(e^{E_F}+1)  \frac{\sqrt{K^2-\Omega^2}}{K^2}\,\theta(K-|\Omega|);  \nonumber
\end{eqnarray}
$\sigma^{\prime}_{xy} = \sigma^{\prime}_{yx} = 0$.
The frequency dispersion of $\sigma^{\prime}_{lm}$, calculated at a fixed value of the wave vector, is shown in  Fig.~2(a) with the solid ($\sigma^{\prime}_{xx}$) and dashed ($\sigma^{\prime}_{yy}$) curves. Note that the real part of DCT is not negative in the whole region of considered frequencies. In the limit of $\Omega \rightarrow K$, the longitudinal component $\sigma^{\prime}_{xx}$ demonstrates a divergence, and the transverse component $\sigma^{\prime}_{yy}$ approaches zero.

\begin{figure}\label{Fig2ab}
\begin{center}
\includegraphics[width=8.5cm]{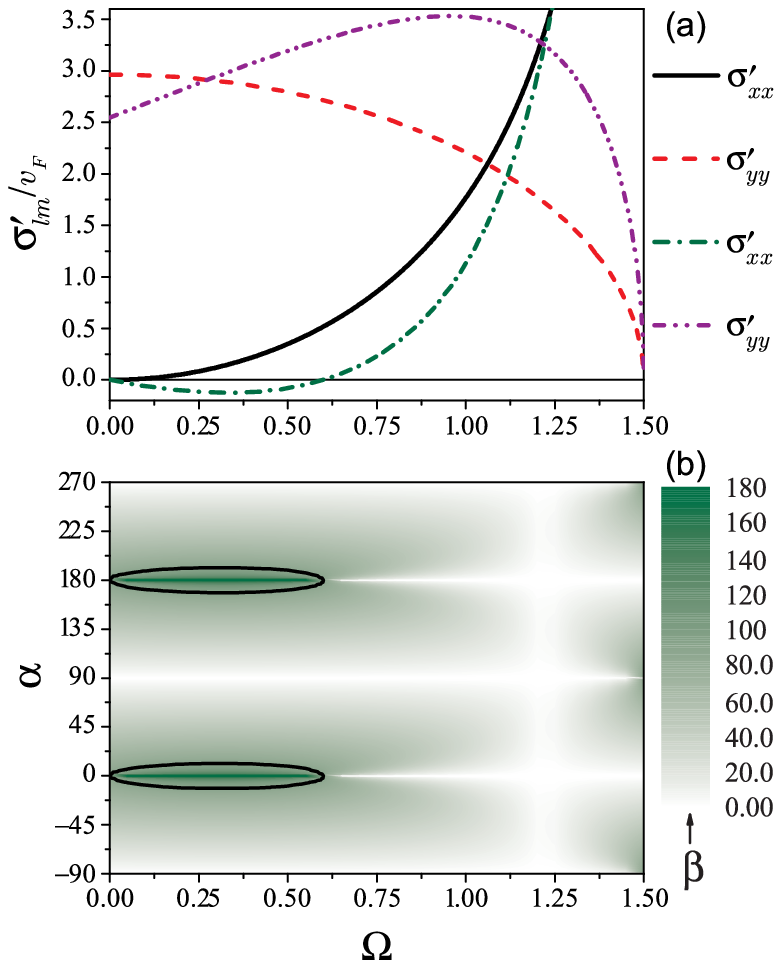}
\caption{(Color online) (a) The real part of normalized DCT, $\sigma^{\prime}_{lm}/v_F$, as a function of frequency (in units of $\omega_0$) calculated at the dimensionless wave vector $K = 1.5$ for $V_d = 0$ (solid and dashed curves) and $V_d = 0.4$ $\,(\gamma = 0^{\circ}$) (dashed-dotted and dashed-double dotted curves). Different curves correspond to different diagonal components of $\sigma_{lm}^{\prime}\,(\sigma_{xy}^{\prime}=\sigma_{yx}^{\prime} = 0)$. (b) Distributions of angle $\beta$ on the plane ($\Omega,\alpha$) corresponding to curves from (a) for $V_d = 0.4$; solid curves ($\beta = 90^\circ$) indicate regions of possible instability.}
\end{center}
\end{figure}

2. {\it In the second case}, the drift velocity is finite ($V_d = 0.4$) and its direction is collinear with the wave vector ${\bf k}$ ($\sin\gamma = 0,  |V_x| = V_d, V_y = 0$). Then we obtain
\begin{eqnarray}\label{resigmak0}
\sigma^{\prime}_{xx} = \frac{v_F \ln{(e^{E_F}+1)}}{\sqrt{K^2-\Omega^2}} \frac{\Omega(\Omega-V_xK)}{(K-V_x\Omega)^2} \, \theta{(K-|\Omega|)},   \\
\sigma^{\prime}_{yy} = v_F \ln{(e^{E_F}+1)} \frac{\sqrt{K^2 - \Omega^2}}{(K - V_x\Omega)^2} \, \theta(K-|\Omega|).  \nonumber
\end{eqnarray}
The calculated frequency dependencies from Eqs.~(\ref{resigmak0}) are shown in Fig. 2(a) with dashed-dotted ($\sigma^{\prime}_{xx}$) and dashed-double dotted ($\sigma^{\prime}_{yy}$) curves.
The frequency behavior of diagonal components at $\Omega \rightarrow K$ is similar to the previous case ($V_d = 0$). However, the occurrence of a finite drift ($V_d = 0.4$) of the electron system adds new features. Namely, if the criterion for the Cherenkov radiation effect is fulfilled $v_d > \omega/k$, i.e., the drift velocity is greater than the ac wave phase velocity, then the longitudinal component $\sigma^{\prime}_{xx} = \sigma^{\prime}_{xx}(\Omega)$ [Fig. 2(a), dashed-dotted curve] changes its sign at the frequency $\Omega = V_xK = 0.65$ ($\nu \simeq 7.5$ THz). It becomes negative in the frequency domain $\Omega < V_xK$. The point of minimum ($\Omega = 0.37$, $\nu \simeq 4.3$ THz) corresponds to maximal increment for the Cherenkov instability in this frequency domain. The transverse component is positive and, as a function of frequency $\sigma^{\prime}_{yy} = \sigma^{\prime}_{yy}(\Omega)$, has a maximum [Fig. 2(a), dashed-double dotted curve].

An important physical quantity associated with the real part of DCT is the power density $P = \langle {\bf j}_1^{\prime}({\bf r},t)\cdot{\bf E}_1^{\prime}({\bf r},t)\rangle$ averaged over the time period $T = 2\pi/\omega$. For monochromatic space and time dependence of the ac field and the current density, the power density $P$ takes the form
\begin{equation} \label{power density}
 P = \frac{1}{2}\left[\sigma_{xx}^{\prime}E_{1x}^{\prime2} + (\sigma_{xy}^{\prime} + \sigma_{yx}^{\prime})E_{1x}^{\prime} E_{1y}^{\prime} + \sigma_{yy}^{\prime}E_{1y}^{\prime2}\right].
\end{equation}
A negative sign of $P$ means that the hot carriers do work over the ac field. In this case, the amplitude of the field of a given wave vector and frequency can be amplified. The occurrence of a $({\bold k},\omega)$-range in which $P$ takes negative values can lead to electrical instability of the considered system. It is evident from Eq.~(\ref{power density}) that $P < 0$ is possible if at least one of the terms in the square brackets is negative and its module value is larger than the sum of others. Alternatively, a positive value of $P$ means dissipation of the ac field energy and stability of the considered system. Formally, considering $P$ as an independent parameter, the equation (\ref{power density}) with $P =$ const can be treated as a family of the second order curves on the plane of variables $(E_{1x}^{\prime},E_{1y}^{\prime})$. These {\it isopower-density curves} are ellipses or hyperbolas depending on sign of the DCT components and $P$. The family of isopower-density curves give a descriptive geometric representation of the DCT as well as the occurrence of possible instability in the considered system. We demonstrate this by letting $\sin\gamma = 0 \, (\sigma_{xy}^{\prime} = \sigma_{yx}^{\prime} = 0)$. Then Eq.~(\ref{power density}) is written in the canonical form
\begin{equation} \label{pd ky0}
 \frac{\sigma_{xx}^{\prime}}{2P} E_{1x}^{\prime2} + \frac{\sigma_{yy}^{\prime}}{2P} E_{1y}^{\prime2} = 1.
\end{equation}
For positive diagonal components ($\sigma_{xx}^{\prime}, \sigma_{yy}^{\prime} > 0$), $P$ also should be positive and
Eq.~(\ref{pd ky0}) specifies a set of ellipses of constant power density. If the signs of $\sigma_{xx}^{\prime}$ and $\sigma_{yy}^{\prime}$ are different, then it is possible to realize the power density of both signs ($P \gtrless 0 $). In this case, the corresponding second order curve is a hyperbola. In particular, $P = 0$ determines asymptotes of the set of hyperbolas, so that the curves corresponding to $P > 0$ and $P < 0$ are located on different sides of these asymptotes. Thus, in going from situation with a positive $P$ to situation with an arbitrary sign of $P$ (i.e., from stability to instability), the set of ellipses is transformed into the set of hyperbolas.

It is worth noting that for a rank-2 symmetrical tensor $\sigma_{lm}^{\prime}$ the dissipative part ${\bf j}_1^{\prime}$ of the current density is normal to the isopower-density curve. This follows from the direct comparison of the normal vector ${\bf n} = \partial P(E_{1x}^{\prime},E_{1y}^{\prime})/\partial {\bf E}_1^{\prime}$,
calculated at a fixed ac electric field $\textstyle{\bf E_{1}^{\prime}}$ with the function $P(E^{\prime}_{1x},E^{\prime}_{1y}) $ given by Eq.~(\ref{power density}), and the vector ${\bf j}_1^{\prime}$.
Thus, each isopower-density curve plotted on the plane $(E_{1x}^{\prime},E_{1y}^{\prime})$ for a given $P$ specifies the field ${\bf E}_{1}^{\prime} = (E_{1x}^{\prime},E_{1y}^{\prime})$ which provides such $P,$ as well as the direction of the current density ${\bf j}_1^{\prime}$ corresponding with the field ${\bf E}_1^{\prime}$. For a general case of $\sin{\gamma} \neq 0$ all components of the DCT are nonzero. In this case, $\sigma_{lm}^{\prime}$ is not a symmetrical tensor and the direction of vector ${\bf j}_1^{\prime}$ does not coincide with the direction of normal vector ${\bf n}$.

Another useful characteristic of induced anisotropy associated with the DCT is the angle $\beta$ between vectors ${\bf j}_1^{\prime}$ and ${\bf E}_1^{\prime}$. It is given by
\begin{equation}\label{anglebeta}
  \cos \beta =  \frac{\sigma_{xx}^{\prime} \cos^2 \alpha + (\sigma_{xy}^{\prime} + \sigma_{yx}^{\prime})\cos\alpha \sin\alpha  + \sigma_{yy}^{\prime}\sin^2 \alpha}
  {[(\sigma_{xx}^{\prime} \cos \alpha + \sigma_{xy}^{\prime} \sin \alpha)^2 + (\sigma_{yx}^{\prime} \cos \alpha + \sigma_{yy}^{\prime} \sin \alpha)^2]^{1/2}} \, ,
\end{equation}
where $\alpha$ is the angle between the wave vector ${\bf k}$ (axis OX) and the ac field ${\bf E}_1^{\prime}$ (Fig.~1). The numerator of the fraction in Eq.~(\ref{anglebeta}) is proportional to the power density $P$. It follows that if the current density ${\bf j}_1^{\prime}$ corresponds to positive $P > 0$, the angle $\beta \in (0, 90^\circ)$; for $P < 0$ the angle $\beta \in (90^\circ, 180^\circ)$. If the power density is zero, then $\beta  = 90^\circ$. In the case of stability, the ac current flows along the ac field under the angle $\beta <90^\circ$; in the case of instability, it flows against the ac field under the angle $\beta >90^\circ$. The angle $\beta$ characterizes the property of the DCT to turn the current density ${\bf j}_1^{\prime}$ relatively to the ac field ${\bf E}_1^{\prime}$. Figure 2(b) shows distributions of $\beta$ on the plane ($\Omega,\alpha$) corresponding to curves with dots from Fig. 2(a). Inside the two regions (bounded with the black curves), which are determined by the condition $\beta = 90^\circ$, the angle $\beta \in (90^\circ,180^\circ]$. Hence, in these regions the considered system may be electrically unstable. Such regions can exist only at frequencies $\Omega < V_x K$. It is also seen that there are regions of very small $\beta$ where the tensor $\sigma_{lm}^{\prime}$ reduces to a scalar. Such representation of the DCT with plots of the angle $\beta$ is very demonstrative; it can be utilized for any other second rank tensor, for example, the dielectric permittivity tensor, etc.

In Fig.~3, we show the curves of a constant power density [Eq.~(\ref{pd ky0})], which correspond to the curves from Fig.~2(a) at $V_d = 0.4$, calculated at different frequency.
The curves which are more distant from the coordinate origin correspond to bigger values of $|P|$. The arrows show vector field of the ac current density [in coordinate axes ($j_{1x}, j_{1y}$)] which direction is normal to the corresponding isopower-density curve. The arrow length reflects relative value of the ac current density.
Figure 3(a) corresponds to $\sigma_{xx}^{\prime} < 0 \, (\Omega=0.3)$; therefore, the curves are hyperbolas. A negative value of $P$ associated with $\sigma_{xx}^{\prime} < 0$ is responsible for the Cherenkov radiation in the region of frequencies $\Omega < KV_d\cos{\gamma}$. Figure 3(b) corresponds to $\sigma_{xx}^{\prime} \approx \sigma_{yy}^{\prime} > 0 \, (\Omega=1.22)$; therefore, the curves in this figure are circles. It is seen from these figures that the angle $\beta$ can substantially be changed with changing the ac field frequency.
\begin{figure}\label{Fig3ab}
\begin{center}
\includegraphics[width=7.5cm]{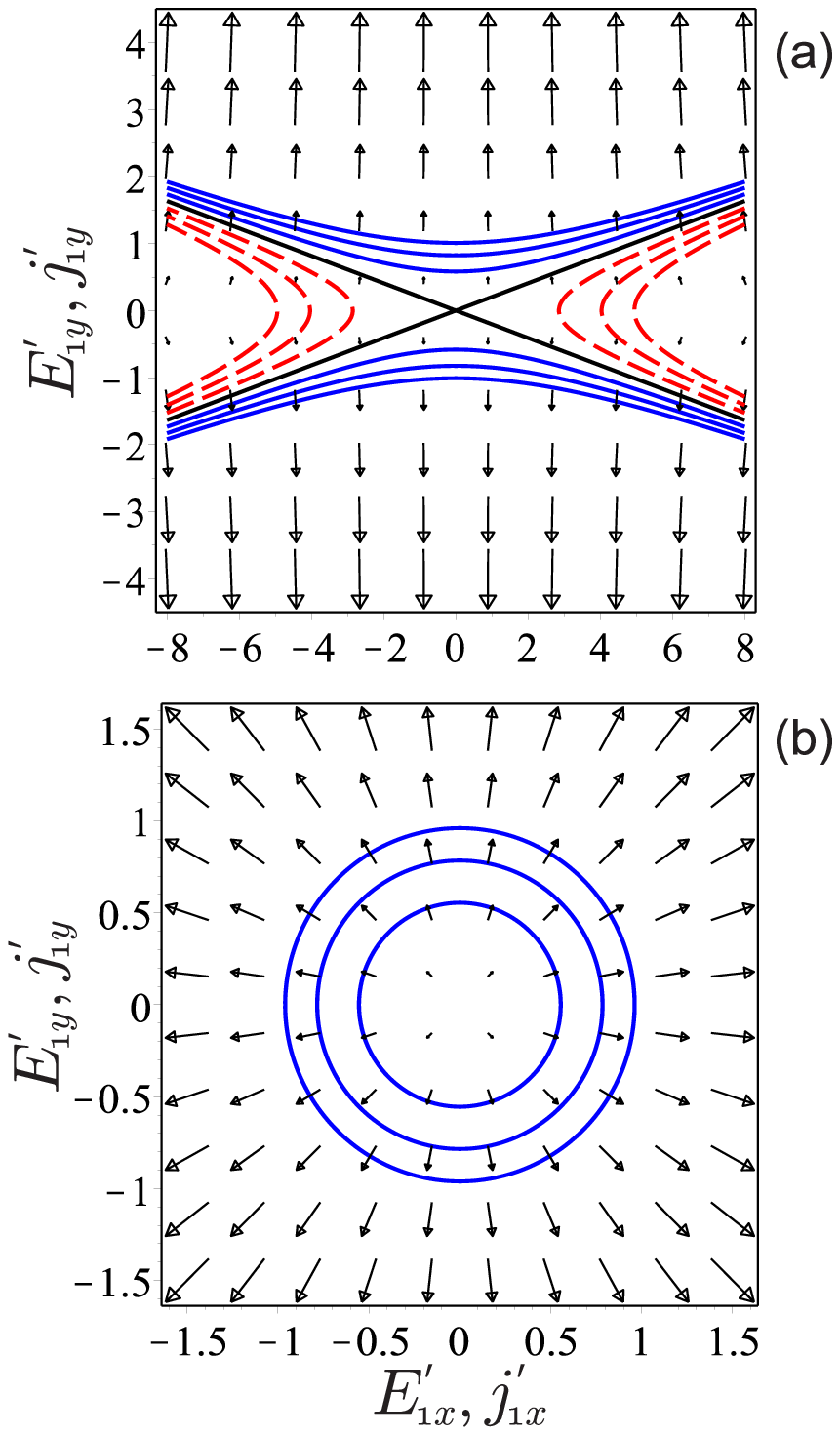}
\end{center}
\caption{(Color online) Lines of constant power density ($P$ = const $\gtrless 0$) corresponding to curves from Fig. 2(a) calculated at different frequency for $V_d = 0.4$: $P > 0$ (solid), $P < 0$ (dashed), and $P = 0$ (straight solid). Arrows represent vector field of the ac current density (relative units). (a) $\Omega = 0.3$ ($\sigma^{\prime}_{xx}<0, \sigma^{\prime}_{yy}>0$) and (b) $\Omega = 1.22$ ($\sigma^{\prime}_{xx} \approx \sigma^{\prime}_{yy} > 0$).}
\end{figure}

3. {\it In the third case}, the drift velocity is finite ($V_d = 0.4$) and its direction is determined by the angle $\gamma = 45^\circ$ (Fig. 1). In this case $V_x = V_y \neq 0$ so that the off-diagonal components of $\sigma_{lm}^{\prime}$ are not a trivial zero. As it was mentioned above, a negative power density ($P < 0$) can be realized if the second term in the square brackets of Eq.~(\ref{power density}) is negative and its module is larger than the rest of terms. This is possible for $(\sigma_{xy}^{\prime} + \sigma_{yx}^{\prime}) < 0$ with $E_{1x}E_{1y} > 0$, or  for $(\sigma_{xy}^{\prime} + \sigma_{yx}^{\prime}) > 0$ with $E_{1x}E_{1y} < 0$. For the former, polarization of the ac field is such that both in-plane components of the field have the same sign, while for the latter they have opposite signs.
In Fig. 4(a), we present the components $\sigma_{lm}^{\prime}$, as well as the sum $(\sigma_{xy}^{\prime} + \sigma_{yx}^{\prime})$, as functions of frequency. In Fig. 4(b), we show distributions of the angle $\beta$. Similar to Fig. 2(b), there are wide regions of frequency ($\Omega < V_x K$), adjoining to the left vertical coordinate axis, where instability associated with the Cherenkov effect can take place. In addition, we revealed narrow regions at high frequencies ($\Omega \rightarrow K$), adjoining to the right vertical coordinate axis, which occurrence is due to a negative sum of the off-diagonal components $\sigma_{lm}^{\prime}$.
\begin{figure}\label{Fig4ab}
\begin{center}
\includegraphics[width=8.5cm]{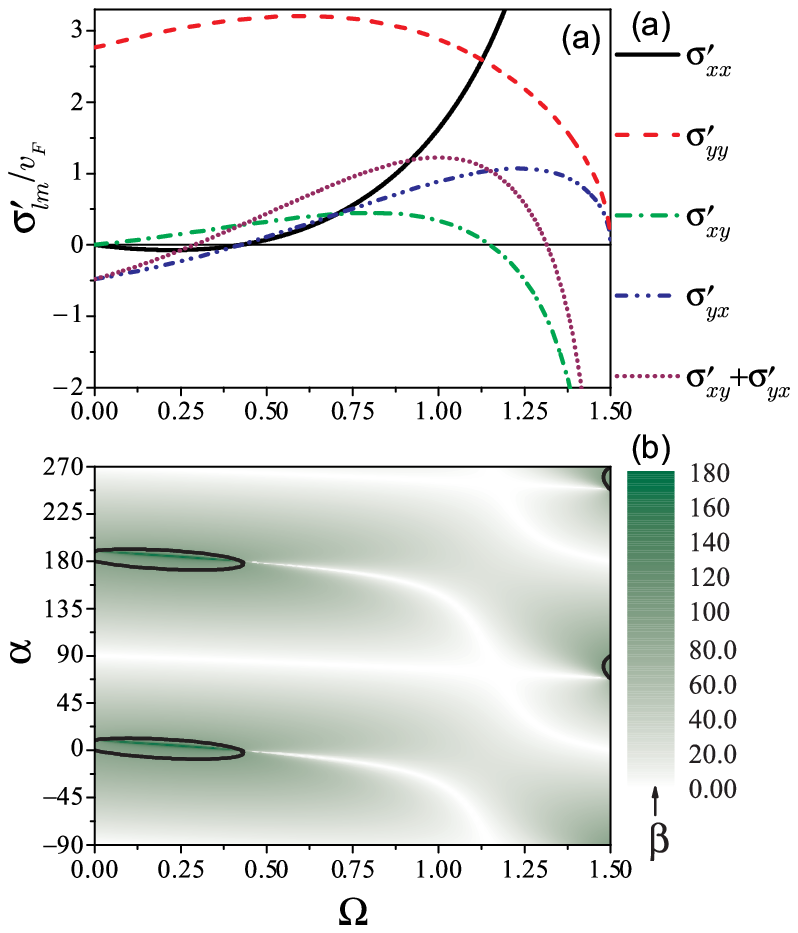}
\caption{(Color online) (a) The real part of normalized DCT, $\sigma^{\prime}_{lm}/v_F$, as a function of frequency (in units of $\omega_0$) calculated at the dimensionless wave vector $K = 1.5 \,(\gamma = 45^{\circ}$) for $V_d = 0.4$. Different curves correspond to different components of $\sigma_{lm}^{\prime}$. (b) Distributions of angle $\beta$ on the plane ($\Omega,\alpha$) corresponding to curves from (a); solid curves ($\beta = 90^\circ$) indicate regions of possible instability.}
\end{center}
\end{figure}

In Fig.~5, we show the isopower-density curves calculated at different frequencies with the same numerical parameters used in Fig. 4(a). The curves correspond to $P > 0$ (solid), $P < 0$ (dashed), and $P = 0$ (straight solid lines). The arrows represent vector field of the ac current density. The curves in Fig. 5(a) were calculated with $\sigma^{\prime}_{xx}<0$ and $(\sigma^{\prime}_{xy}+\sigma^{\prime}_{yx}) < 0$ at $\Omega = 0.3$. The curves in Fig. 5(b) were calculated with $\sigma^{\prime}_{xx}>0$ and $(\sigma^{\prime}_{xy}+\sigma^{\prime}_{yx}) < 0$ at $\Omega = 1.49$, i.e., at $\Omega \rightarrow K$.
\begin{figure}\label{Fig5ab}
\begin{center}
\includegraphics[width=7.5cm]{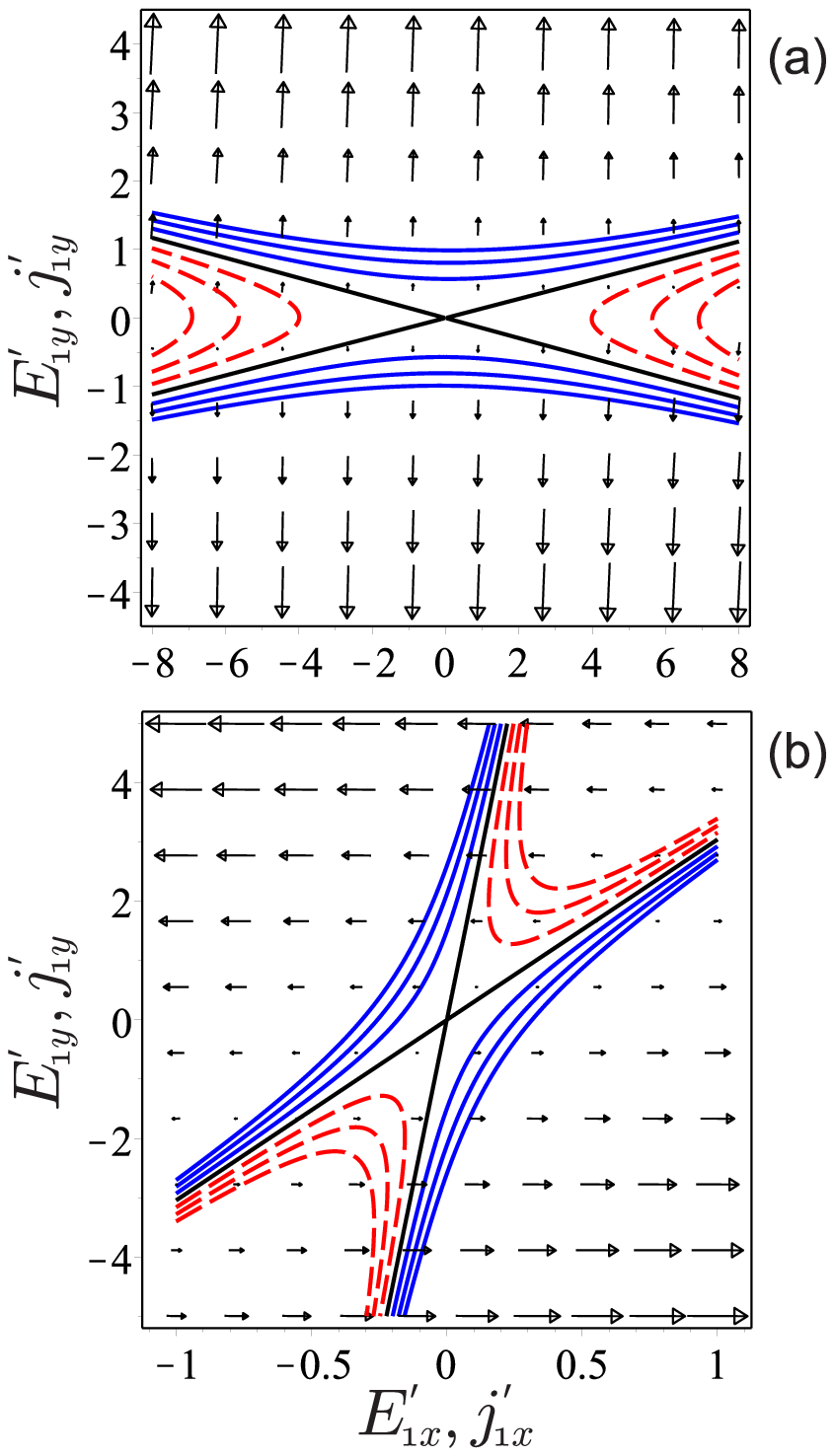}
\caption{(Color online) Lines of constant power density ($P$ = const $\gtrless 0$) corresponding to curves from Fig. 4(a) calculated at different frequency: $P > 0$ (solid), $P < 0$ (dashed), and $P = 0$ (straight solid). Arrows represent vector field of the ac current density (relative units). (a) $\Omega = 0.3, \, \sigma^{\prime}_{xx}<0, \sigma^{\prime}_{xy}+\sigma^{\prime}_{yx}>0$ and (b) $\Omega = 1.49 \,, \sigma^{\prime}_{xx}>0, \sigma^{\prime}_{xy}+\sigma^{\prime}_{yx}<0$.}
\end{center}
\end{figure}

4. {\it In the forth case}, the drift velocity is finite ($V_d = 0.4$) and its direction is perpendicular to the ${\bf k}$ vector, $\gamma = 90^\circ$ (Fig. 1). In this case $V_y = V_d$ and $V_x = 0$ which excludes the possibility for the Cherenkov instability in the considered system.
To illustrate more clearly the influence of the off-diagonal term ($\sigma^{\prime}_{xy}+\sigma^{\prime}_{yx}$) on the sign of $P$, we compare all nonzero components $\sigma^{\prime}_{lm}$ calculated as a function of frequency in Fig. 6(a). It follows that $P < 0$ may be realized due to the term ($\sigma^{\prime}_{xy}+\sigma^{\prime}_{yx}$) which is the only negative at high frequency ($\Omega \rightarrow K$); the rest of the terms in Eq.~(\ref{power density}) are definite positives at the considered frequencies. The corresponding distributions of angle $\beta$ are presented in Fig. 6(b), which demonstrates at what frequency the negative power density can take place. It is seen that the regions of possible instability ($P<0$) are slightly wider compared to the similar regions in Fig. 4(b).
\begin{figure}\label{Fig6ab}
\begin{center}
\includegraphics[width=8.5cm]{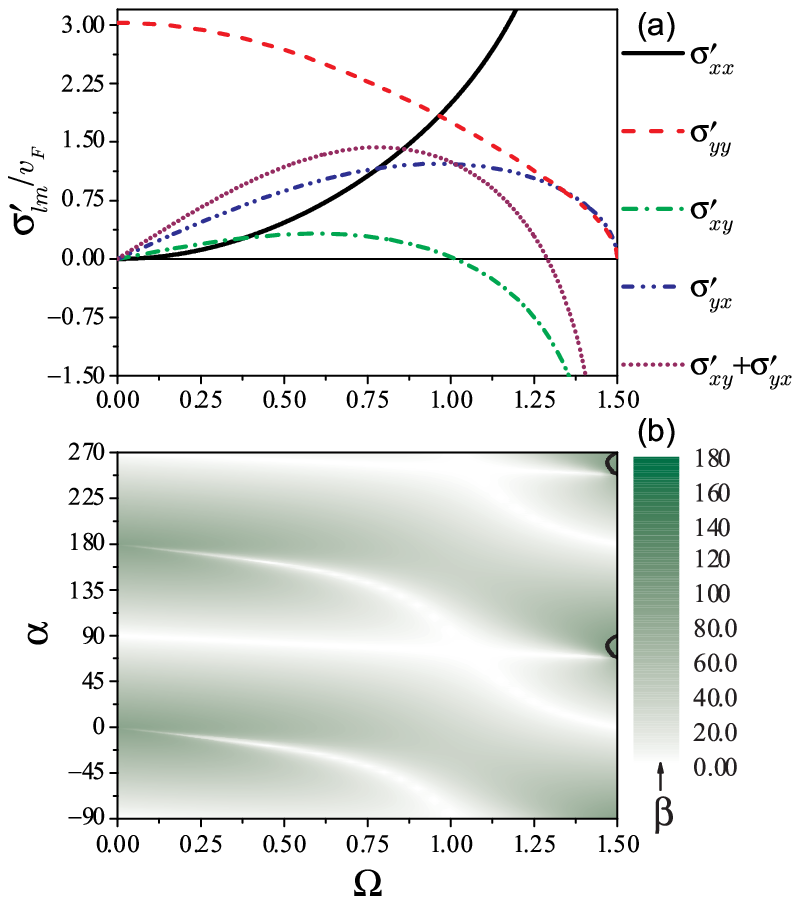}
\caption{(Color online) (a) The real part of normalized DCT, $\sigma^{\prime}_{lm}/v_F$, as a function of frequency (in units of $\omega_0$) calculated at the dimensionless wave vector $K = 1.5 \, (\gamma = 90^{\circ}$) for $V_d = 0.4$. Different curves correspond to different components of $\sigma_{lm}^{\prime}$. (b) Distributions of angle $\beta$ on the plane ($\Omega,\alpha$) corresponding to curves from (a).}
\end{center}
\end{figure}

In Fig.~7, we show the isopower-density curves corresponding to curves from Fig.~6(a) calculated at different frequencies. The curves demonstrate a qualitatively different behavior. At $\Omega = 0.3$, we get $P>0$ as both $\sigma^{\prime}_{xx} > 0$ and $\sigma^{\prime}_{xy}+\sigma^{\prime}_{yx} > 0$ [Fig.~7(a)].   At $\Omega = 1.48$, we can get $P\gtrless0$ as $\sigma^{\prime}_{xx} > 0$ but $(\sigma^{\prime}_{xy}+\sigma^{\prime}_{yx}) < 0$ [Fig. 7(b)]. The results show that with  $\gamma =90^\circ$ a negative power density is achieved in the case when the Cherenkov criterion is not fulfilled, i.e., due to the negative off-diagonal term alone $(\sigma^{\prime}_{xy}+\sigma^{\prime}_{yx}) < 0$. We have also revealed that the term $(\sigma^{\prime}_{xy}+\sigma^{\prime}_{yx})$ may become negative at $\Omega \rightarrow K$ in a wide region of the angle $\gamma \in(0^\circ, 180^\circ)$.
\begin{figure}\label{Fig7ab}
\begin{center}
\includegraphics[width=7.5cm]{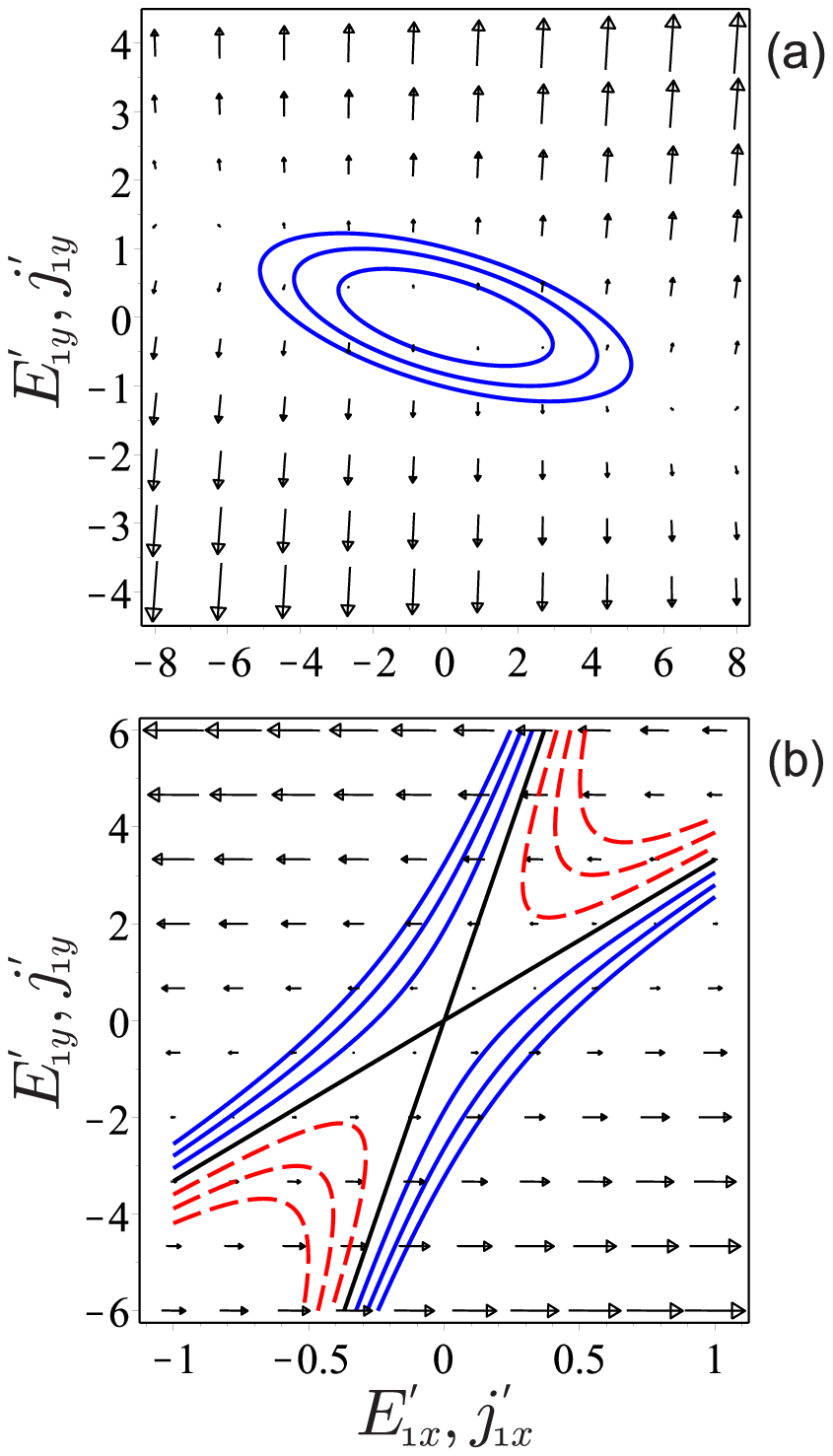}
\caption{(Color online) Lines of constant power density ($P$ = const $\gtrless 0$) corresponding to curves from Fig. 6(a) calculated at different frequency. (a) $\Omega = 0.3$ ($P > 0$) and (b) $\Omega = 1.48$ ($P \gtrless 0$). $P > 0$ (solid), $P < 0$ ( dashed), and $P = 0$ (straight solid). Arrows represent vector field of the ac current density (relative units).}
\end{center}
\end{figure}

Thus, there exists a certain set of the specific parameter values at which the negative power density ($P<0$) may be realized. This comprises three different cases with $\sigma^{\prime}_{xx}<0$ in the Cherenkov region of frequencies (Fig.~2); with $(\sigma^{\prime}_{xy}+\sigma^{\prime}_{yx}) < 0$ in the limit of high frequencies $\Omega \rightarrow K$ (Fig.~6); and with both $\sigma^{\prime}_{xx}<0$ and $(\sigma^{\prime}_{xy}+\sigma^{\prime}_{yx}) < 0$ simultaneously (Fig. 4). We also found that if $\gamma\in(180^\circ, 360^\circ)$, then one may achieve $P < 0$ at the positive off-diagonal term $(\sigma^{\prime}_{xy}+\sigma^{\prime}_{yx}) > 0$ due to the ac field polarization with $E_{1x}^{\prime}E_{1y}^{\prime} < 0$ in the limit of $\Omega \rightarrow K$.

To conclude this section, we note that so far we considered the ac field of arbitrary configuration given by the two vectors ${\bf E}_{1}^{\prime}$ and ${\bf k}$, where ${\bf E}_{1}^{\prime}$ was not necessarily parallel to the wave vector ${\bf k}$. Let us briefly discuss the case when the effective ac field is of an electrostatic origin. Then, the Fourier components of the field and the electrostatic potential are connected by ${\bf E}_1^{\omega{\bf k}} = -i{\bf k}\varphi_1^{\,\omega{\bf k}}$.  In this case, ${\bf E}_{1}^{\prime}$ is collinear with ${\bf k}$ (Fig. 1), and we get for the power density
\begin{equation} \label{pdelectrostat}
 P = \frac{1}{2}\sigma_{xx}^{\prime} k^2 |\varphi_1^{\,\omega{\bf k}}|^2  \,,
\end{equation}
which now replaces  (\ref{power density}). From the comparison of (\ref{pdelectrostat}) and (\ref{power density}), we conclude that in the case of ac electrostatic field the negative power density ($P<0$) can be realized only on the Cherenkov criterion $\Omega < KV_d\cos\gamma$.

\section{Imaginary part of dynamic conductivity}

In this section, the frequency and wave vector behavior of the imaginary part of DCT is analyzed by separating $\sigma_{lm}^{\prime\prime}$ in Eq.~(\ref{sigma exact}) (Appendix). In the analysis, we will use the same exemplary cases of particular parameter values considered in the previous section. In the absence of the electron drift (case 1, $V_d = 0$) and for small space dispersion (${\bf k} \rightarrow 0$), the DCT reduces to pure imaginary scalar $\sigma_{lm} = i\sigma^{\prime\prime}\delta_{lm}$
with
\begin{equation}\label{ImsigmaK0Vd0}
 \sigma^{\prime\prime}({\bf k}=0,\omega) = \frac{e^2 k_B T_e}{\pi\hbar^2\omega} \ln(e^{\varepsilon_F/k_BT_e}+1)    \,.
\end{equation}
Hence, the dependence of $\sigma^{\prime\prime}({\bf k}=0,\omega)$ on frequency is similar to the Drude-Lorentz conductivity at high frequencies ($\omega\tau_p \gg 1$). A similar result was obtained for intraband contribution into $\sigma^{\prime\prime}({\bf k},\omega)$ in Ref.~\onlinecite{Falkovsky_2007} within a quantum approach for low electric field and at finite lattice temperature $T$ ($\hbar\omega < k_BT$), when the electron distribution over the energy is equilibrium one [Eq.~(10) in Ref.~\onlinecite{Falkovsky_2007}]. The expression (\ref{ImsigmaK0Vd0}) extends this result to the region of hot electrons in high electric fields, when the electron temperature $T_e$ is higher than the lattice temperature. If space dispersion is large, then the imaginary part of DCT reads
\begin{eqnarray}\label{imsigmaVd0}
\sigma^{\prime\prime}_{xx} =- \frac{v_F \ln{(e^{E_F}+1)}}{K^2}\Omega\left(1-\frac{\Omega\,\theta{(|\Omega|-K)}}{\sqrt{\Omega^2-K^2}} \right),   \\
\sigma^{\prime\prime}_{yy} = \frac{v_F \ln{(e^{E_F}+1)}}{K^2}\left(\Omega-\sqrt{ \Omega^2-K^2} \, \theta(|\Omega|-K)\right).  \nonumber
\end{eqnarray}
From these expressions, we see that the diagonal elements $\sigma^{\prime\prime}_{xx}$ and $\sigma^{\prime\prime}_{yy}$ are linear functions of $\Omega$ at $\Omega < K$ and take nonzero values at $\Omega > K$. In the limit $\Omega\rightarrow\infty$, we obtain the asymptotic behavior $\sigma^{\prime\prime}_{xx} = \sigma^{\prime\prime}_{yy} = v_F \ln{(e^{E_F}+1)}/2\Omega \sim \Omega^{-1}$. At $\Omega = K$, the function $\sigma^{\prime\prime}_{xx}(\Omega)$ has an infinite discontinuity, whereas $\sigma^{\prime\prime}_{yy}(\Omega)$ is a continuous function of $\Omega$.
\begin{figure}\label{Fig8ab}
\begin{center}
\includegraphics[width=8.5cm]{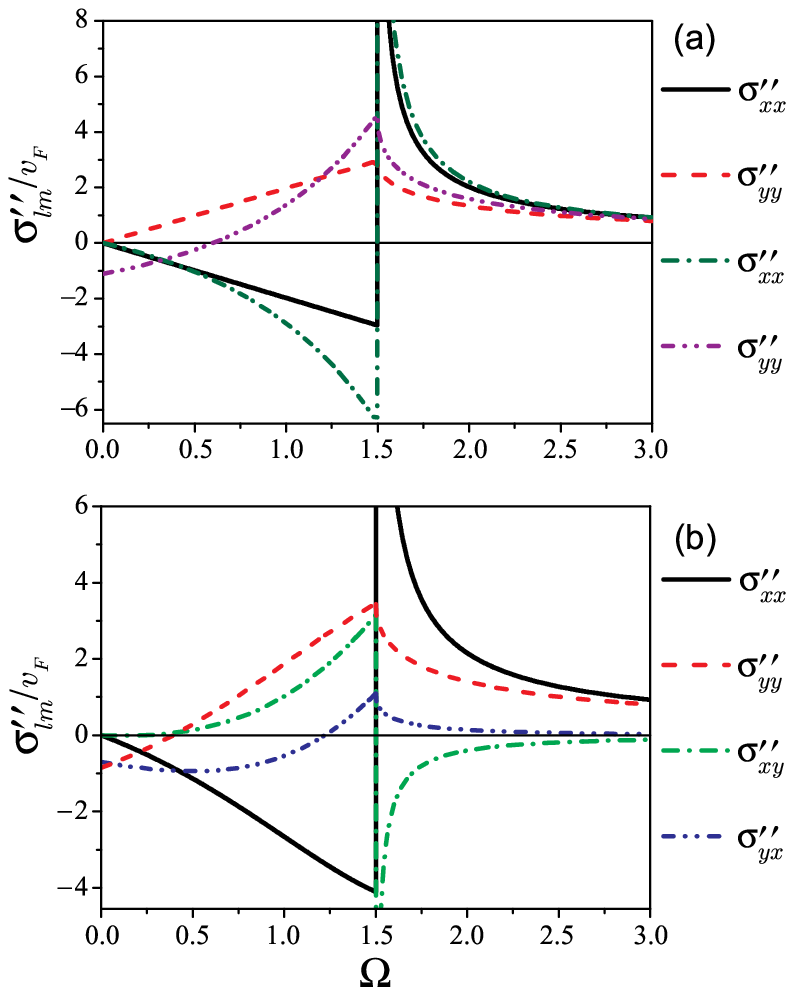}
\caption{(Color online) The imaginary part of normalized DCT, $\sigma^{\prime\prime}_{lm}/v_F$, as a function of frequency (in units of $\omega_0$) calculated at different values of the dimensionless wave vector. (a)  $K = 1.5 \, (\gamma = 0^\circ$), $\sigma_{xy}^{\prime\prime}=\sigma_{yx}^{\prime\prime} = 0$; $V_d = 0$ (solid and dashed curves) and $V_d = 0.4$ (dashed-dotted and dashed-double dotted curves). (b) $K = 1.5 \, (\gamma = 45^\circ$), $V_d = 0.4$. Different curves correspond to different components of $\sigma_{lm}^{\prime\prime}$.}
\end{center}
\end{figure}

In a general case, the corresponding integrals are evaluated by the theory of residues (see the Appendix). The results of numerical calculations of the imaginary part of DCT as a function of frequency are shown in Fig.~8. The parameter values used in the calculations are for Fig. 8(a) the same as for Fig. 2(a) [case 1 ($V_d = 0$) and case 2 ($\sin\gamma = 0, V_d = 0.4$)]; for Fig. 8(b) the same as for Fig. 4(a) [case 3 ($\gamma = 45^\circ, V_d = 0.4$)]. As mentioned above, the off-diagonal elements of $\sigma_{kl}$ are zero for the considered cases 1 and 2. The diagonal elements $\sigma^{\prime\prime}_{k=l}$, unlike the real part $\sigma^{\prime}_{k=l}$, are not zero at $\Omega > K$. It is seen from Fig.~8(a) that the asymptotic behavior of $\sigma^{\prime\prime}_{xx}$ (and $\sigma^{\prime\prime}_{yy}$) at $\Omega \rightarrow K$ and $\Omega \rightarrow \infty$ for a finite electron drift are similar to the case of $V_d = 0$. The occurrence of the electron drift leads to an increase in $|\sigma^{\prime\prime}_{k=l}|$ near the point $\Omega = K$. The transversal diagonal component $\sigma^{\prime\prime}_{yy}$ takes negative values in a frequency region $\Omega <K$, starting from zero frequency. For comparison, in Fig. 8(b), we show the results obtained for $\gamma = 45^\circ$, in which case the off-diagonal components $\sigma^{\prime\prime}_{xy}$ and $\sigma^{\prime\prime}_{yx}$ are not zero. The frequency behavior of the diagonal components $\sigma^{\prime\prime}_{k=l}$ is similar to that shown in Fig. 8(a) for $V_d \neq 0$. The component $\sigma^{\prime\prime}_{xy}$ has an infinite discontinuity, approaching $-\infty$ when $\Omega \rightarrow K+0$. The frequency behavior of the off-diagonal component $\sigma^{\prime\prime}_{yx}$ is similar to that of the diagonal component $\sigma^{\prime\prime}_{yy}$.

\section{Role of energy band nonlinearity}

So far, our calculations were restricted to the Dirac-spectrum approximation. In the integral of
Eq.~(\ref{sigma lm}), the integration is over all possible values of the electron momentum ${\bf p}$. Therefore, corrections to the Dirac spectrum may be important at certain values of the model parameters. In this section, we address the question of how the frequency and wave vector behavior of the dynamic conductivity can be changed by including corrections to the Dirac cone spectrum. Recently, the dynamic and optical properties in graphene have been studied taking into account the nonlinear energy dispersion.~\cite{Stauber,Yang} The full energy band spectrum has been obtained in the tight-binding approximation.~\cite{Wallace,Neto} In using such the energy spectrum in Eqs.~(\ref{f0}) and (\ref{sigma lm}), the DCT can be calculated numerically performing the integration in Eq.~(\ref{sigma lm}) over the entire Brillouin zone. For the purpose of this section, it is sufficient to use a quadratic correction to the linear term $\varepsilon({\bf p}) = v_Fp$ which follows from Taylor expansion of the full band spectrum close to the Dirac point,
\begin{equation}\label{nonlin}
 \varepsilon({\bf p}) = v_Fp - \frac{3E_t a^2}{8\hbar^2} \sin(3\vartheta) p^2 \,.
\end{equation}
Here $E_t \simeq$ 2.8 eV is the nearest-neighbor hopping energy, $a \simeq 1.42 {\AA}$ is the carbon-carbon distance, and $\vartheta$ is the angle in momentum space.~\cite{Neto} Assuming the strongest nonlinearity [letting $\sin(3\vartheta) = 1$] and using the dimensionless variables $P = p/p_0$, $p_0 = k_BT_e/v_F$, the spectrum (\ref{nonlin}) can be rewritten as $E_p \equiv \varepsilon({\bf p})/k_B T_e = P - (1/2)\alpha_0 P^2$, where $\alpha_0 = 3k_BT_eE_ta^2/4v_F^2\hbar^2$ is a dimensionless parameter. For example, if we take $T_e = 400$ K, then the numerical estimation gives $\alpha_0 = 3.6\times10^{-3} \ll 1$. Substituting $E_p$ in Eq.~(\ref{sigma lm}), we find
\begin{equation}\label{sigma lm_P}
  \sigma_{lm} = - i \frac{e^2 gv_F}{(2\pi\hbar)^2} \int \frac{d^2 p}{\omega - v_F {\bf k}\cdot{\bf p}/p
  + \alpha_0 v_F {\bf k}\cdot{\bf p}/p_0 + i\delta}
  \frac{p_l}{p} \frac{\partial f_0}{\partial p_m}  \,.
\end{equation}
As in Sec.~III, the integration is convenient to be performed in the polar coordinate system. Importantly, the denominator of the integrand in Eq.~(\ref{sigma lm_P}) contains not only the polar angle but also the electron momentum, which modifies essentially the pole of the integrand compared to the Dirac cone approximation. As a consequence, the diagonal component $\sigma_{xx}^{\prime}$ does not go to infinity at the limit $\omega\rightarrow v_F k$. Instead, it takes a finite hight and width at the resonance frequency. The distribution function in Eq.~(\ref{f0}) now depends on the parameter $\alpha_0$ through the modified electron spectrum,  $f_0 = f_0({\bf p}, \alpha_0)$. Note that $f_0({\bf p}, \alpha_0)$ cannot be normalized according to Eq.~(\ref{epsilon F}), because it does not approach zero when the momentum goes to infinity. Nevertheless, it is possible to evaluate an asymptotic value for the integral in Eq.~(\ref{sigma lm_P}), utilizing the small parameter $\alpha_0$ and expanding the function $f_0({\bf p}, \alpha_0)$ into a Taylor series
$f_0({\bf p}, \alpha_0) = f_0({\bf p}) + \alpha_0 [\partial f_0({\bf p}, \alpha_0)/\partial \alpha_0]_{\alpha_0=0} + \cdots$. Substituting this distribution function in Eq.~(\ref{sigma lm_P}), we obtain the asymptotic expansion for the DCT $\sigma_{lm} = \sigma_{lm}^{0} + \alpha_0
\sigma_{lm}^{1} + \cdots$. Further, we retain only the main (zeroth order) term  $\sigma_{lm}^{0}$ and ignore all higher orders in $\alpha_0$ of this expansion.  Then, the expression for $\sigma_{lm}^{0}$ is given in Eq.~(\ref{sigma lm_P}) with the function $f_0({\bf p})$ given in Eq.~(\ref{f0}). The real and imaginary parts of the DCT, $\sigma_{lm} \simeq \sigma_{lm}^{0}$, can be derived in a fashion similar to Sec.~III, where they have been calculated within the Dirac cone approximation. As the derived analytical formulas are rather complicated, we demonstrate only the results of numerical computation and compare them with those obtained in the Dirac cone approximation. 
Using the polar coordinates ($P,\theta$), the longitudinal component $\sigma_{xx}^{\prime}$ can be written as
\begin{equation}\label{sigma lm_0 prime}
  \sigma_{xx}^{\prime} = - \frac{v_F}{2 K} \int_0^{\infty} \frac{dP}{1 - \alpha_0 P} \int_0^{2\pi} d\theta \left( P \frac{\partial f_0}{\partial P}\cos^2 \theta - \cos\theta \sin\theta \frac{\partial f_0}{\partial \theta} \right) \delta (\tilde{s} - \cos\theta)  \,,
\end{equation}
where $\tilde{s} = s/(1 - \alpha_0 P)$. Due to the delta function $\delta (\tilde{s} - \cos\theta)$, the angle integral in (\ref{sigma lm_0 prime}) is not zero if only $|\tilde{s}| \leq 1$. Then, we have
\begin{equation}\label{sigma lm_0 prime1}
  \sigma_{xx}^{\prime} = \frac{v_F}{2 K} \int_0^{\infty} dP \frac{P I_{xx}}{1 - \alpha_0 P} \theta(1 - |\tilde{s}|) = \frac{v_F}{2 K} \int_0^{P_c} dP \frac{P I_{xx}}{1 - \alpha_0 P}  \,,
\end{equation}
where $P_c = (1 - |\Omega|/K)/\alpha_0$,
\begin{equation}\label{Ixx}
 I_{xx} = \frac{\tilde{s}^2 - V_x \tilde{s}}{(1 - \tilde{s}^2)^{1/2}} [f_0^{+}(1 - f_0^{+}) + f_0^{-}(1 - f_0^{-})]   \,,
\end{equation}
and $f_0^{\pm} = \{\exp[(1 - V_x \tilde{s} \mp V_y \sqrt{1 - \tilde{s}^2})P - E_F]  + 1\}^{-1}$.
In particular, for the case of $V_d = 0$, Eq.~(\ref{sigma lm_0 prime1}) becomes
\begin{equation}\label{sigma xx_0}
  \sigma_{xx}^{\prime} =
  - v_F \frac{\Omega^2}{K^3} \int_0^{P_c} \frac{P dP}{(1 - \alpha_0 P)^{2} \sqrt{(1 - \alpha_0 P)^2 - \Omega^2 /K^2}} \frac{\partial f_0}{\partial P}  \,.
\end{equation}

\begin{figure}\label{Fig9ab}
\begin{center}
\includegraphics[width=8.5cm]{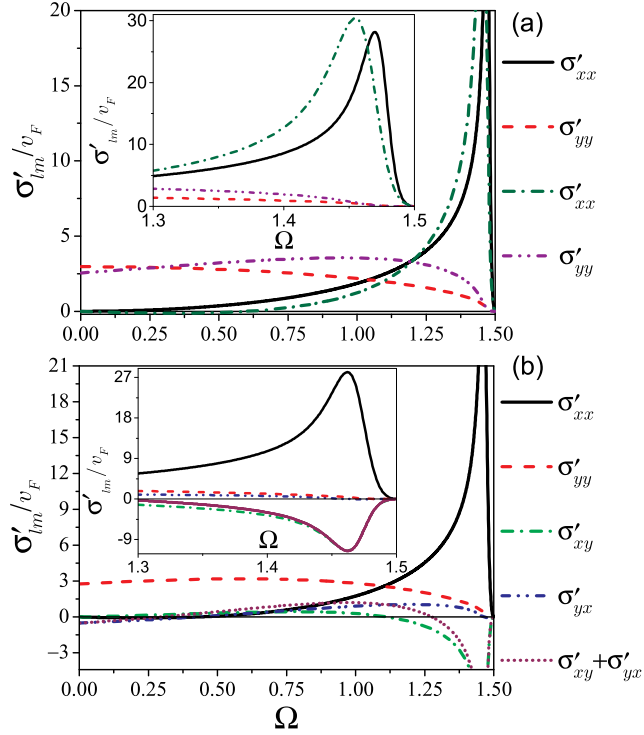}
\caption{(Color online)  The real part of normalized DCT, $\sigma^{\prime}_{lm}/v_F$, calculated with the quadratic correction to the graphene Dirac spectrum as a function of frequency (in units of $\omega_0$). (a) $K = 1.5 \, (\gamma = 0^\circ$), $\sigma_{xy}^{\prime\prime}=\sigma_{yx}^{\prime\prime} = 0$;  $V_d = 0$ (solid and dashed curves) and $V_d = 0.4$ (dashed-dotted and dashed-double dotted curves). (b) $K = 1.5 \,(\gamma = 45^\circ$),  $V_d = 0.4$. Different curves correspond to different components of $\sigma_{lm}^{\prime}$. The inserts show details of the resonance.}
\end{center}
\end{figure}
Figure~9 shows the frequency dependence of the real part of DCT calculated with the quadratic correction to the linear energy spectrum. The parameter values used for Fig.~9(a) are the same as for Fig. 2(a), and for Fig. 9(b) are the same as for Fig. 4(a). As expected, the divergence at the resonance frequency ($\Omega \rightarrow K$) is removed due to the adopted quadratic correction. The integrals in Eqs.~(\ref{sigma lm_0 prime1}) and (\ref{sigma xx_0}) approach zero in the limit $P_c \rightarrow 0$ ($|\Omega| \rightarrow K$). Thereby, we observe a normal resonance curve with the finite peak and width values. In Fig. 9(a), the peak values are $\sigma_{xx}^{\prime}/v_F = 28.3$ at $\Omega = 1.47$ ($V_d = 0$) and $\sigma_{xx}^{\prime}/v_F = 31.0$ at $\Omega = 1.45$ ($V_d = 0.4$). In Fig. 9(b), the peak values are $\sigma_{xx}^{\prime}/v_F = 28.2$ (maximum) at $\Omega = 1.46$ and $\sigma_{xy}^{\prime}/v_F = -11.52$ (minimum) at the same $\Omega$ [inserts in Fig. 9(a) and 8(b)]. Far from the resonance, the real part of DCT practically coincides with that obtained in the previous section but remains finite near the resonance. Note that the off-diagonal element $\sigma_{xy}^{\prime}$ changes its sign from positive (for small frequencies) to negative (near the resonance frequency). The element $\sigma_{yx}^{\prime}$ changes its sign from negative (for small frequencies) to positive and, again, to negative (near the resonance frequency). As a consequence, the sum $(\sigma_{xy}^{\prime}+\sigma_{yx}^{\prime})$ shows two frequency windows in which it takes negative values [dotted curve in Fig. 9(b)]. Within these windows, the considered graphene system can be electrically unstable, and an external electromagnetic wave of such frequency can be amplified.

\section{Screening effect}

Within the employed theoretical approach, the screening effect can be considered as follows. The weak ac electric field ${\bf E}_1({\bf r},t)$ induces a modulation of the electron density $n_1({\bf r},t)$ in the graphene layer. In turn, the induced electron charge $-en_1({\bf r},t)$ generates an electrostatic potential $\varphi_1(z,{\bf r},t)$ and an additional lateral electric field ${\bf F}_1({\bf r},t) = -\partial \varphi_1(z,{\bf r},t)/\partial{\bf r}|_{z=0}$, which changes the initial field ${\bf E}_1({\bf r},t)$ and thereby gives rise to screening effect. Further, the electric field ${\bf F}_1({\bf r},t)$ is included in the expression for the ac current density $j_{1l} = \sigma_{lm} (E_{1m} + F_{1m})$ (summation over the repetitive indices). Since the field ${\bf F}_{1}$ is proportional to ${\bf E}_{1}$, the ac current density can be written as $j_{1l} = \sigma^{s}_{lm} E_{1m}$, where the renormalized DCT $\sigma^{s}_{lm}$ takes onto account the effect of screening.

The electrostatic potential is given by the Poisson equation
\begin{equation}\label{potential}
  \left(\frac{\partial^2}{\partial z^2} + \frac{\partial^2}{\partial {\bf r}^2}\right) \varphi_1(z,{\bf r}) =  \frac{4 \pi e}{\varepsilon_0} n_1({\bf r})\delta(z) \,,
\end{equation}
which solution is
\begin{equation}\label{potential 1}
 \varphi_1(z,{\bf r},t) = - \frac{2 \pi e}{\varepsilon_0 k} n_1^{\omega {\bf k}}\, \exp(-k|z|)\, \exp[i({\bf k}\cdot{\bf r} - \omega t)  \,.
\end{equation}
Then, it follows that the induced lateral electric field is along the wave vector ${\bf k}$,
\begin{equation}\label{F 1}
 {\bf F}_1 ({\bf r},t) = i{\bf k} \frac{2 \pi e}{\varepsilon_0 k} n_1^{\omega {\bf k}}\, \exp[i({\bf k}\cdot{\bf r} - \omega t)  \,.
\end{equation}
The local change of the electron density $n_1({\bf r},t)$ can be found from the continuity equation
\begin{equation}\label{continuity}
  \frac{\partial n_1({\bf r},t)}{\partial t} - \frac{1}{e} \frac{\partial  {\bf j}_1 ({\bf r},t)}{\partial {\bf r}}   = 0  \,,
\end{equation}
which takes the form of an equation for the Fourier coefficient $n_1^{\omega {\bf k}}$
\begin{equation}\label{continuity F}
  i \omega n_1^{\omega {\bf k}} + \frac{i}{e}k_l\sigma_{lm}\left(E_{1m}^{\omega {\bf k}} + \frac{2\pi i e}{\varepsilon_0 k}k_m n_1^{\omega {\bf k}}\right) =0  \,.
\end{equation}
Then, we obtain
\begin{equation}\label{density 1}
  n_1({\bf r},t) = - \frac{k_l \sigma_{lm}E_{1m}({\bf r},t)}{e (\omega + i \tau^{-1})}
\end{equation}
and
\begin{equation}\label{field F}
 {\bf F} _1({\bf r},t) = - \frac{2\pi i}{\varepsilon_0} \frac{\bf k}{k}
  \frac{k_l \sigma_{lm} E_{1m}({\bf r},t)}{\omega + i \tau^{-1}}        \,,
\end{equation}
where the quantity $\tau = \varepsilon_0 k/2\pi k_l \sigma_{lm} k_m$ has the dimensionality of time and depends on the wave vector. Upon substituting ${\bf F} _1$ from Eq.~(\ref{field F}) into the current density $j_{1l} = \sigma_{lm} (E_{1m} + F_{1m}) = \sigma^{s}_{lm} E_{1m}$ and comparing both the expressions, one finally obtains
\begin{equation}\label{sigma s}
 \sigma_{xx}^s = \sigma_{xx} -
  \frac{2\pi i}{\varepsilon_0 k} \frac{k_l \sigma_{lx}\sigma_{xm}k_m}{\omega + i \tau^{-1}}    \nonumber  \,,
\end{equation}
\begin{equation}
 \sigma_{yy}^s = \sigma_{yy} -
  \frac{2\pi i}{\varepsilon_0 k} \frac{k_l \sigma_{ly}\sigma_{ym}k_m}{\omega + i \tau^{-1}}  \,,
\end{equation}
\begin{equation}
 \sigma_{xy}^s = \sigma_{xy} -
  \frac{2\pi i}{\varepsilon_0 k} \frac{k_l \sigma_{ly}\sigma_{xm}k_m}{\omega + i \tau^{-1}}     \nonumber  \,,
\end{equation}
\begin{equation}
 \sigma_{yx}^s = \sigma_{yx} -
  \frac{2\pi i}{\varepsilon_0 k} \frac{k_l \sigma_{lx}\sigma_{ym}k_m}{\omega + i \tau^{-1}}     \nonumber    \,.
\end{equation}
The result \eqref{sigma s} is quite general and is valid not only for the graphene, but also for an arbitrary quantum well system.

The equations \eqref{sigma s} are simplified in the chosen coordinate system: $k_x = k$, $k_y = 0$ (Fig. 1).
As an example, we consider the wave propagating along the dc field ($\gamma = 0^\circ$). Then Eqs. (\ref{sigma s}) read
\begin{equation}\label{sigma xx s ky0}
  \sigma_{xx}^s = \frac{\sigma_{xx}}{1 + i (\omega \tau)^{-1}} \,,
\end{equation}
$\sigma_{yy}^s = \sigma_{yy}$, and  $\sigma_{xy}^s = \sigma_{yx}^s = 0$. The similar expression is obtained for the total electric field ${\bf E}_t = {\bf E}_1 + {\bf F}_1$ which longitudinal component is $E_{tx} = E_{1x}/[1 + i(\omega \tau)^{-1}]$, and the transverse component is not screened $E_{ty} = E_{1y}$. The denominator in these expressions can be treated as an effective dielectric function $\varepsilon(k, \omega) = 1 + i(\omega \tau)^{-1}$, where $\tau^{-1} = (2\pi/\varepsilon_0)k\sigma_{xx}$. Such renormalization of the DCT changes its frequency behavior near the resonance. Indeed, considering the limit $\Omega \rightarrow K$, we obtain from Eq.~(\ref{resigmak0}) the asymptotic expression
\begin{equation}\label{sigma xx Re}
  \sigma_{xx}^{\prime} = v_F \frac{\ln(e^{E_F}+1)}{1 - V_x} \frac{1}{(2K)^{1/2}
  (K - \Omega)^{1/2}} \,,
\end{equation}
which diverges at the resonance as $\sigma_{xx}^{\prime}|_{\Omega \rightarrow K} \sim (K - \Omega)^{-1/2}$. Separating the real and imaginary parts in Eq.~(\ref{sigma xx s ky0}), we find
\begin{equation}\label{sigma xx s Re}
  (\sigma_{xx}^s)^{\prime} = \frac{\sigma_{xx}^{\prime}}{A^2(k,\omega) + B^2(k,\omega)} \,,
\end{equation}
where $A(k,\omega) = (1 - 2\pi k \sigma_{xx}^{\prime\prime}/\varepsilon_0 \omega)$ and $B(k,\omega) = 2\pi k \sigma_{xx}^{\prime}/\varepsilon_0 \omega$. Taking into account that at the considered limit $\sigma_{xx}^{\prime\prime}$ is finite and $\sigma_{xx}^{\prime}$ goes to infinity, then $B(k,\omega) \sim \sigma_{xx}^{\prime} \rightarrow \infty$, we can write according to (\ref{sigma xx s Re}) $(\sigma_{xx}^s)^{\prime} \simeq (\varepsilon_0 v_F/2\pi)^2 (1/\sigma_{xx}^{\prime})$. Now, utilizing Eq.~(\ref{sigma xx Re}), we obtain the asymptotic expression
\begin{equation}\label{sigma xx s Re As}
 (\sigma_{xx}^s)^{\prime}|_{\Omega \rightarrow K} = v_F \frac{\varepsilon_0^2 (1 - V_x)}{4\pi^2 \ln(e^{E_F}+1)}
  (2K)^{1/2}(K - \Omega)^{1/2} \,,
\end{equation}
in which $(\sigma_{xx}^s)^{\prime} \sim (K - \Omega)^{1/2}$, i.e., it approaches zero for $\Omega \rightarrow K$.
The total electric field $E_{tx}(x,t) = (1/2)[E_{tx}^{\omega {\bf k}}\exp(ikx-i\omega t) + c.c.] = |E_{tx}^{\omega {\bf k}}|\cos(kx-\omega t + \varphi_0)$ is characterized by the similar behavior. Specifically, we find the relative amplitude $\tilde{E}_{tx}^{\omega {\bf k}} = |E_{tx}^{\omega {\bf k}}/E_{1x}^{\omega {\bf k}}|$ of the total field
\begin{equation}\label{relatampl}
 \tilde{E}_x^{t\omega {\bf k}} = \Big[A^2(k,\omega) + B^2(k,\omega) \Big]^{-1/2}
\end{equation}
and the phase shift $\cos \varphi_0 = A(k,\omega)/[A^2(k,\omega) + B^2(k,\omega)]^{1/2}$. From Eq.~(\ref{relatampl}), we obtain the asymptotic behavior near the resonance $\tilde{E}_{tx}^{\omega {\bf k}} \sim (\sigma^{\prime}_{xx})^{-1} \sim (K - \Omega)^{1/2} \rightarrow 0$. We may conclude that due to self-consistent electrostatic potential the considered divergence is removed leading to finite values of the DCT at the resonance.

\section{Conclusion}

We have investigated the dynamic conductivity tensor, $\sigma_{lm}({\bf k}, \omega)$, of doped graphene subjected to a strong dc electric field. The steady-state transport has been characterized in terms of the hot Dirac quasiparticles with the electron (hole) temperature $T_{e}$ and the drift velocity $v_{d}$. The analysis has been carried out in the most general fashion considering a weak ac field of arbitrary configuration and regarding the ac field vector and the wave vector ${\bf k}$ as independent vectors. We found that specific features of dispersion of DCT are essentially determined by the Dirac energy spectrum and by the existing resonance at $\omega = v_Fk$, at which the phase velocity of the wave in the direction of propagation coincides with the carrier velocity.

The real part of DCT takes nonzero values (except the points where it changes its sign) at frequencies below the resonance ($\omega < v_Fk$) and diverges [($\sigma_{xx}({\bf k}, \omega$)] at $\omega \rightarrow v_Fk$. This means dissipation of the energy of ac wave since in relation to such carriers the ac field is effectively a steady-state one and therefore can do work, which is not zero when averaging over the wave period. We have shown that deviation from the Dirac-like spectrum at high energies and screening by the charge carriers regularizes the divergence, such that the resulting resonance curve is characterized by definite (finite) peak and width values. The imaginary part of DCT is nonzero for all considered frequencies and goes to zero for $\omega \rightarrow \infty$.
The anisotropy of the carrier distribution function induced by a strong dc electric field leads to a new isolated direction (the electron drift) in addition to the wave vector ${\bf k}$, so that tensor structure of the DCT is complicated. We have revealed certain ac field configurations for which the ac power density is negative. This has allowed us to indicate regions of terahertz frequency for possible electrical (drift) instability in the considered graphene system. We suggest that the knowledge of the DCT dispersion controlled by a strong dc electric field may be useful for further progress in graphene plasmonics and applications.

Finally, we note that the theoretical formalism developed in this work is quite general and may be useful for calculations of spatially dispersive dynamic response under a strong dc electric field in graphene-analogous 2D semiconductor nanomaterials, such as transition-metal dichalcogenides, the silicon and germanium counterparts of graphene, etc.~\cite{Xu,Chhowalla,Tang} In addition, it would also be interesting to explore, for comparison, other steady-state regimes of the high-field transport with known distribution function of the hot carriers, in particular, obtained with advanced numerical methods for the graphene\cite{Li1,Li2} and graphene-like nanomaterials\cite{Li3} in combination with the results obtained in this work.

\section*{ACKNOWLEDGMENTS}

This work was partially supported by NASU (State Targeted Scientific-Technical Program $``$Nanotechnology and Nanomaterials$"$ 2010-2014, No. 2.3.4.22), STCU (Targeted R\&D Initiatives Program 2012-2014, No. 5716), and by a grant for young scientists of NAS of Ukraine. The work performed at North Carolina State University was supported, in part, by FAME (one of six centers of STARnet, a SRC program sponsored by MARCO and DARPA).

\newpage
\appendix
\section{Evaluation of the kinematic integrals}\label{App-A}

The kinematic integrals in Eq.~\eqref{sigma lm} are evaluated by integrating over ${\bf p}$ in the polar coordinate system $p_x = p \cos\theta$ and $p_y = p \sin\theta$, in which the direction of the $x$ axis in the momentum space is chosen to be along the ${\bf k}$ vector, $\theta$ is the polar angle, and $d^2p = p\, dp\,d\theta$. With the function $f_0({\bf p})$ of Eq.~(\ref{f0}), the integral over $p$ from $0$ to $\infty$ is easily evaluated, and we get
\begin{eqnarray} \label{IntPoAlpha}
 \sigma_{xx} = i\frac{v_F\ln{(e^{E_F}+1)}}{2\pi K} \int\limits_0^{2\pi}{\!\!d\theta \frac{\cos^2{\theta}-V_x\cos\theta}{(s-\cos\theta)(1-V_x \cos\theta - V_y \sin\theta)^2}}, \\ \nonumber
 \sigma_{yy} = i\frac{v_F\ln{(e^{E_F}+1)}}{2\pi K} \int\limits_0^{2\pi}{\!\!d\theta \frac{\sin^2{\theta}-V_y\sin\theta}{(s-\cos\theta)(1-V_x\cos{\theta}-V_y\sin{\theta})^2}}, \\ \nonumber
 \sigma_{xy} = i\frac{v_F\ln{(e^{E_F}+1)}}{2\pi K} \int\limits_0^{2\pi}{\!\!d\theta \frac{\sin{\theta}\cos{\theta}-V_y\cos\theta}{(s-\cos\theta)(1-V_x\cos{\theta}-V_y\sin{\theta})^2}}, \\ \nonumber
 \sigma_{yx} = i\frac{v_F\ln{(e^{E_F}+1)}}{2\pi K} \int\limits_0^{2\pi}{\!\!d\theta \frac{\sin{\theta}\cos{\theta}-V_x\sin\theta}{(s-\cos\theta)(1-V_x\cos{\theta}-V_y\sin{\theta})^2}},
\end{eqnarray}
where $\textstyle{V_x=V_d\cos{\gamma}}$ and $\textstyle{V_y=V_d\sin{\gamma}}$. Here, we use the dimensionless variables introduced in Sec. III. The complex quantity $\textstyle{s=(\Omega+i\delta)/K}$ is reduced to the dimensionless phase velocity $\Omega/K$ of the propagating wave in the limit of $\delta \rightarrow 0$. Further, we make a substitution $z = \exp(i\theta)$, then $\cos\theta = (z + z^{-1})/2$ and $\sin\theta = (z - z^{-1})/2i$; the integrals over the angle $\theta$, from $0$ to $2\pi$, become the integrals of rational functions of $z$ in the complex plane, where the integration is over a unit circle $|z|=1$ centered in the origin. These functions (the integrands) are given by
\begin{eqnarray} \label{Flm}
 F_{xx}(z) = \frac{-(z^2+1)(z^2-2V_x z+1)}{(z-z^+)(z-z^-)(z-z_+)^2(z-z_-)^2}, \\ \nonumber
 F_{yy}(z) = \frac{(z^2-1)(z^2-2 i V_y z-1)}{(z-z^+)(z-z^-)(z-z_+)^2(z-z_-)^2}, \\ \nonumber
 F_{xy}(z) = \frac{i(z^2+1)(z^2-2 i V_y z-1)}{(z-z^+)(z-z^-)(z-z_+)^2(z-z_-)^2}, \\ \nonumber
 F_{yx}(z) = \frac{i(z^2-1)(z^2-2V_x z+1)}{(z-z^+)(z-z^-)(z-z_+)^2(z-z_-)^2},
\end{eqnarray}
where $\textstyle{z^{\pm}=s\pm\sqrt{s^2-1}}$ and $\textstyle{z_{\pm}=(1\pm\sqrt{1-V_d^2})/(V_x-iV_y)}$ are the poles of the functions $F_{lm}(z)$ of the first and second order, respectively. It is a straightforward procedure to evaluate the integrals by using the theorem of residues, which gives each integral of the functions (\ref{Flm}) equals the sum of residues at a number of isolated poles located inside the integration contour, multiplied by $2\pi i$. The residue at an $n$-order pole at $z=a$ is evaluated by~\cite{Morse}
\begin{equation} \label{Res}
 Res^{(n)}F_{lm}(a) = \frac{1}{(n-1)!}\lim_{z\to a} \frac{d^{\,(n-1)}}{dz^{\,(n-1)}}\Bigl[(z-a)^nF_{lm}(z)\Bigr].
\end{equation}

It should be noted that the expressions for the first order poles $z^{\pm}$ contain a square root of complex function, which is a double-valued function in the complex plane. If we chose a cut along the real axis in the complex plane $s$ from $-1$ to $+1$, then only one pole $z = z^{-}$ is inside the integration path over the unit circle. The reason is that, at such cut, the function $z^{-}$ conformally maps points of the plane $s$ into the interior of the unit circle; correspondingly, the function $z^{+}$ transforms them onto the exterior of the unit circle. The poles $z_{\pm}$ do not depend on frequency, and there is only one pole $z_{-}$ inside the unit circle. Thus, finally, the two poles $z = z^{-}$ and $z = z_{-}$ are inside the unit circle, and we get the exact expression
\begin{equation} \label{sigma exact}
 \sigma_{lm} = i\frac{2v_F \ln{(e^{E_F}+1)}}{K(V_x-iV_y)^2}\left[Res^{(1)}F_{lm}(z^-)+Res^{(2)}F_{lm}(z_-)\right].
\end{equation}
Since the drift velocity $V_d$ is in the denominator of formula \eqref{sigma exact}, the situation with the absence of the electron drift should be treated as the limit of $V_d \rightarrow 0$.

\end{document}